\documentstyle[onecolumn]{mn}

\def\be{\begin{equation}}
\def\ee{\end{equation}}
\def\sg{\sigma}
\def\mm{\langle \log M \rangle}
\def\mmo{\langle \log M \rangle_{outer}}
\def\mmi{\langle \log M \rangle_{inner}}
\begin{document}
\title{Evolution of the mass function of the Galactic globular
cluster system}

\author[E.Vesperini]{E.Vesperini\thanks{E-mail:vesperin@falcon.phast.umass.edu
}
\\ Department 
of Physics and Astronomy, University of Massachusetts, Amherst, MA, 01003, USA}
\maketitle
\begin{abstract}
In this work we investigate the evolution of the mass function of the
Galactic globular cluster system (GCMF) taking into account the effects of
stellar evolution, two-body relaxation, disk shocking and  dynamical
friction on the evolution of individual globular clusters.
We have adopted a log-normal initial GCMF and  considered a wide range
of initial values for the dispersion, $\sigma$, 
 and the mean value, $\langle \log M\rangle$.
We have studied in detail the dependence on the initial
conditions of the final values of $\sigma$, $\mm$, of the fraction of
the initial number of clusters surviving after one Hubble time, and of
the difference between the properties of the GCMF of clusters closer
to the Galactic center and the properties of those located in the
outer regions of the 
Galaxy.  In most of the cases considered evolutionary
processes alter significantly the initial population of globular
clusters and the disruption of a significant number of globular
clusters leads to a flattening in the spatial distribution of
clusters in the central regions of the Galaxy.  The initial log-normal
shape of the GCMF  is preserved in most cases and if a power-law in
$M$ is adopted 
for the initial GCMF, evolutionary processes tend to modify it into a
log-normal GCMF. The difference between  initial and final values
of $\sg$ and $\mm$ as well as the difference between  the final values of
these parameters for inner and outer clusters can be positive or negative
depending on initial conditions. A significant effect of evolutionary
processes does not necessarily give rise to a strong
trend of $\mm$ with the galactocentric distance.
The existence of a particular initial GCMF able to keep its initial
shape and parameters unaltered during the entire evolution through a
subtle balance between disruption of clusters and evolution of the
masses of those which survive, suggested in Vesperini (1997), is confirmed.
\end{abstract}
\section{Introduction}
Investigation of the luminosity function of the globular cluster
system (hereafter GCLF) of our Galaxy and of globular cluster systems
in external galaxies has been the subject of 
many observational works (see e.g. Secker 1992, McLaughlin 1994,
Abraham \& van den Bergh 1995, Kissler-Patig 1997,
 Harris 1991 and references therein) 
because of its relevance
for a number of issues of great astrophysical interest, such as the
formation of 
globular clusters, the role of the external galactic field on the
evolution of their properties and the possibility of using the
turnover of the GCLF
as a standard candle calibrated on the values of globular cluster
systems in the Local Group for the determination of the distance of
external galaxies (see e.g. Jacoby et al. 1992). 

The determination of the distances of external galaxies by means of
the turnover of the GCLF relies on
the assumed constancy of the properties of the GCLF for galaxies of different
structure and type which is quite surprising: 
in fact, unless one advocates  a  scenario
in which clusters in different galaxies have different initial
conditions and different dynamical histories but all leading to the
same final state, such common characteristics imply that the process of
formation of globular clusters does not depend on the galactic
environment and that evolutionary processes (tidal stripping, disk
and bulge shocking, dynamical friction; see e.g. Meylan \& Heggie
1997, for a recent review on the  dynamics of globular clusters) have
not played a  relevant role in 
determining the present  properties of globular clusters.

 While the
present knowledge of the 
processes leading to the formation of globular clusters is still very
uncertain (see e.g. Fall \& Rees 1985, Harris \& Pudritz 1994, Vietri
\& Pesce 1995, Elmegreen \& Efremov 1997) many theoretical
investigations (Aguilar, Hut \& 
Ostriker 1988, Chernoff, Kochanek \& Shapiro 1986, Chernoff \& Shapiro
1987, Vesperini 1994, 1997, Okazaki \& Tosa 1995, Murali \& Weinberg
1997, Gnedin \& 
Ostriker 1997, Baumgardt 1998) have
clearly shown that  evolutionary processes should have altered
significantly the 
initial properties of globular clusters in galaxies like the
Milky Way by causing the complete disruption of a fraction of them and
by altering the initial properties of the surviving ones.
As it has been  shown in several
works (see e.g. Caputo \& Castellani 1984, Chernoff, Kochanek \&
Shapiro 1986, Vesperini 1994, 1997, Murali \& Weinberg 1997, 
Ostriker \& Gnedin 1997, Baumgardt 1998) the inner 
regions of the Galaxy are 
those where evolutionary effects are expected to be more efficient and
where to look for traces of their effects. Indeed Chernoff
\& Djorgovski (1989) have provided the first observational evidence of
this by showing that the fraction of clusters in the post-core
collapse phase increases as the distance from the Galactic center
decreases. In a subsequent work Bellazzini et al. (1996) have shown
the existence for clusters located in the inner regions of the Galaxy
of a significant correlation between concentration and 
galactocentric distance in the sense of more concentrated clusters
being on average closer to the Galactic center. This is
 likely to result from evolution (Vesperini
1994, 1997, Bellazzini et al. 1996) occuring faster for clusters
located in the inner
regions of the Galaxy than for those  in the outer ones.

The situation concerning differences between the GCLF of inner and
outer clusters is far from being clear. In a recent work Gnedin (1997)
 has carried out an analysis of the available observational data for
clusters in the Galaxy, M31 and M87, and his results seem to point to
the existence of
some differences between the GCLF of inner and outer clusters for all
these three  galaxies: inner clusters tend to be brighter and to
have smaller dispersions than outer clusters.
Kavelaars \& Hanes (1997) have  addressed the same issue for the Milky Way and
M31 and their conclusion is that, while there is no significant
difference in the 
mean luminosity of inner and outer clusters, their distributions are
actually different, the inner clusters being well described by a
Gaussian in the magnitude with a dispersion significantly smaller than
that of outer clusters; as for M87, Harris et al. (1998) in a recent
analysis have not found any significant radial gradient in the
properties of the GCLF for clusters with masses $M>10^5 M_{\odot}$
except for a possible trend for the dispersion of the GCLF of the
innermost region of M87 that they have considered to be smaller than
the dispersion of the GCLF of clusters in the outer regions of the
galaxy. 
As discussed in Gnedin (1997) the reason for
the difference between  his analysis and that by Kavelaars and
Hanes could reside in the different  
statistical methods adopted for deriving the parameters of the
distribution.  
A trend for inner clusters to be brighter than the outer ones
was previously suggested by  van den Bergh (1995) and Crampton et al.
(1985) for clusters in the Milky Way and in M31 respectively.

From a theoretical point of view, as we said above, this trend
is consistent with that expected to result from evolutionary
processes, at least for some initial GCMFs,
which are more efficient in the inner regions of the Galaxy 
where they can efficiently  disrupt low-mass clusters (see e.g. fig.
2 in Vesperini 1997). In Vesperini (1994, 1997) the evolution of the
properties of a system of globular clusters located in a model of the
Milky Way under the effects of relaxation, disk shocking and, in an
approximate way, of dynamical
friction, starting from three different initial GCMF, has been
investigated. In all these cases 
a trend for inner clusters to be more massive than outer
clusters was obtained as a result of evolutionary processes, with the
extent of the difference depending on the initial GCMF chosen.

While in Vesperini (1994,1997), besides addressing some general issues
on the evolution and the properties of the GCMF, we investigated the
origin of some observed correlations between structural properties of
individual globular clusters and between structural parameters of
clusters and their position inside the host galaxy, in this work we
will focus our attention on and investigate in larger detail the
evolution of the properties of the GCMF of a globular cluster system
located in a model for the Milky Way adopting some analytical formulae
for the time 
evolution of the masses of individual clusters obtained by the results
of a large set of $N$-body simulations carried out by Vesperini \&
Heggie (1997).

 We will
adopt a  log-normal distribution  for the initial GCMF and we will consider a
wide range of different initial conditions largely
spanning the space of the initial parameters (dispersion and mean) 
of the GCMF. Different functional forms for
the initial GCMF have also been studied to investigate  the evolution
of their shape  and in particular to establish if the current gaussian
shape could result from an initial GCMF with a different functional form.

We will devote a section to the comparison of our results
with observational data, but we point out that due 
to some assumptions made in our analysis, which will be discussed
in sect.2 together with the description of the method adopted for our
investigation, an exact comparison of our results with the available
observational data is beyond the scope of our work. The main goal of
our analysis is that of  providing general indications
on the evolution of the properties of the GCMF, of the spatial distribution and
 the fraction of the initial number of clusters surviving after one
Hubble time. The evolution of the shape of the GCMF for the whole
sample of clusters and the possible development of 
differences between the GCMF of clusters located in the inner and in
the outer regions of the Galaxy will be thoroughly investigated paying
particular attention to the dependence of the final results on the
initial conditions.  
The issue, raised in Vesperini (1997), of the  possible existence
of  a dynamical  ``equilibrium'' GCMF able to preserve its initial shape and
parameters for one Hubble
time through a subtle balance between disruption of clusters and
evolution of the masses of the surviving ones is further investigated.

The scheme of the paper is the following. In sect.2 the method adopted
for our study is described; in section 3 
we report the main results of the investigation: after a preliminary
qualitative discussion on the evolution of the GCMF in section 3.1, 
in sections 3.2-3.6  
we describe the results obtained not including
the effects of disk shocking; in particular section 3.2 discusses the
possible evolutionary paths  of  the parameters of the GCMF depending
on the initial conditions (see e.g. figure 3) and the existence of a
GCMF of dynamical equilibrium is shown and discussed in detail (see e.g.
figure 4c), 
section 3.3 is focussed on the dependence of  the final  GCMF on the distance
from the Galactic center (see e.g. figure 10 and 11), in section 3.4
the time evolution of some systems is followed in detail and some
other aspects of  the GCMF able to stay in dynamical equilibrium are
studied (see figure 13). Section 3.5 and 3.6 are devoted to the
study of the fraction of surviving clusters and their spatial
distribution in the Galaxy respectively; in section 3.7 we discuss the
results obtained including the effects of disk shocking.
In section 4 we describe the results obtained assuming a power-law
initial GCMF and section 5 is 
devoted to the comparison of our results with observational data.
Summary and conclusions are in section 6.
\section{Method}
In order to calculate the evolution of the GCMF we need to know
the  time evolution of the masses of individual globular clusters in
the system.  
In this work we will adopt the analytical formulae obtained in Vesperini \&
Heggie (1997)  which supply the mass at any time $t$ of a cluster with
initial mass 
$M_i$ and moving in a circular orbit at a distance
$R_g$ from the Galactic center.  These have been obtained by fitting
the results 
of a large set of
$N$-body simulations following the evolution of globular clusters driven by
internal relaxation, stellar evolution, disk shocking and including the effects
of the tidal field of the Galaxy. 
We summarize here for convenience the main assumptions made in the
simulations carried out by Vesperini \& Heggie (1997) and we refer to
that paper for further details.
\begin{enumerate}
\item Clusters are assumed to move on  circular orbits in a Keplerian
potential determined by a point mass $M_g$ equal to the mass of the Galaxy
inside the adopted galactocentric distance $R_g$. For the simulations including
the effects of disk shocking, it is assumed that  orbits cross the galactic
disk perpendicularly. The circular speed has been taken equal to
$v_c=220$ km/s.
\item Disk shocking has been included according to the model described in
Chernoff, Kochanek \& Shapiro (1986) and the same two-component disk model
obtained by Chernoff et al. by a fit of the Bahcall's (1984) determination of
acceleration in the solar neighbourhood has been adopted. This is an
exponential isothermal disk 
model with  scale heights equal to 175 pc and 550 pc and  scale length $h=3.5
$ Kpc.
\item An initial multi-mass King model with $W_0=7$ has been
adopted. A set of simulations starting with $W_0=5$ has been also
carried out in  Vesperini \& Heggie (1997) and it was shown that the
evolution of the total mass does not depend strongly on the initial
concentration of the cluster. 
\item The initial stellar mass function has been taken equal to a power-law
$dN(m)=m^{-2.5}dm$ between $0.1m_{\odot}$ and $15m_{\odot}$.
\item Stellar evolution is modelled following the same model used in Chernoff
\& Weinberg (1990) and the mass lost by each star is assumed to escape
immediately from the cluster.
\end{enumerate}

Fitting the results of N-body simulations not including the effects of
disk shocking,  Vesperini
\& Heggie (1997) have obtained the following expression for the time
evolution of 
the total mass of a cluster with initial mass $M_i$ and galactocentric distance
$R_g$  
\be
{M(t) \over M_i}=1-{\Delta M_{st.ev.}\over M_i}-{0.828\over F_{cw}}t \label{mev}
\ee
where $t$ is  time measured in Myr, ${\Delta M_{st.ev.}\over M_i}$ is the
mass loss due to stellar evolution (see eq.[10] in Vesperini \& Heggie
1997) and 
$F_{cw}$ is a parameter, introduced by Chernoff \& Weinberg (1990),
which is 
proportional to the relaxation time and defined as
\be
F_{cw}\equiv{M_i\over M_{\odot}}{R_g\over Kpc}{1\over \ln(N)}{220 \hbox{km
s}^{-1} \over v_c}
\ee
where $M_i$ and $N$ are, respectively,  the initial mass and the initial number
of stars in the cluster, $R_g$ is the distance from the Galactic
center and $v_c$ the circular velocity around the Galaxy.

From the simulations including disk shocking an expression analogous to
eq.(\ref{mev}) has been derived but  with the factor $0.828/F_{cw}$ replaced by
the following factor $\lambda$ (see Vesperini \& Heggie 1997 for
further details 
on the derivation)
\be
\log \lambda=0.6931-1.46\log R_g-1.134\log F_{cw}+0.2916\log F_{cw} \log R_g.
\ee

A comparison of the characteristic cluster lifetime from
eq.(\ref{mev}) with other estimates present in the literature has been
made in Vesperini \& Heggie (1997). More recently in an analysis of
the evolution of globular cluster systems Baumgardt (1998) has adopted
a formula for the mass loss introduced by Wielen (1988) but with a new
value for the numerical factor present in the original expression
obtained by fitting some numerical results present in the literature;
for the initial conditions considered in Vesperini \& Heggie (1997)
the cluster lifetime obtained by Baumgardt turns out to be 10 per cent
longer than that obtained by eq. (\ref{mev}) without considering the
mass loss associated to stellar evolution (stellar evolution is not
considered in Baumgardt's analysis) while if we include the effects of
stellar evolution in the estimate of the cluster lifetime derived from
eq. (\ref{mev}), the lifetime obtained by Baumgardt is 30 per cent
longer than ours.

As we have anticipated in the Introduction, in  this investigation  many
different initial conditions for the GCMF of the system have been
investigated. For
any system considered, $10^4$ random values of $M_i$ according to the
chosen  initial 
GCMF have been calculated and a random value for the galactocentric distance
from a distribution such that the number of cluster per cubic Kpc is
proportional to $R_g^{-3.5}$ has been assigned to each cluster with
$1 \hbox{ Kpc}<R_g<20\hbox{ Kpc}$. This functional form for the  number density
profile is suggested by what is observed in our Galaxy between
$4\hbox{ Kpc}<R_g<20 
\hbox{ Kpc}$ where the sample of observed clusters is likely to be
complete  (see e.g. Zinn 1985).

Once the initial conditions have been set, the GCMF at any time $t$
can be easily  
calculated by means of eq. (\ref{mev}). In order to consider the
effects of dynamical friction, clusters whose timescale of orbital
decay (see e.g. Binney \& Tremaine 1987) is
less than $t$ are removed from the sample at that time.

Our investigation will be focussed on initial conditions characterized by a
log-normal initial GCMF; as we will see in the following sections in most cases
the gaussian shape is well preserved during the entire evolution until
the final 
state at $t=15 $ Gyr. The median value, $\langle \log M\rangle$, and
the non-parametric estimate of the dispersion
 
\be
\sigma=0.7415(Q_{75}-Q_{25}),
\ee
where $Q_{75}$ and $Q_{25}$ are the 75th and 25th percentiles
 of the distribution from  the final sample of surviving clusters,
are the parameters used throughout the paper in order to characterize the
final GCMF.

\section{Results}
\subsection{Preliminary remarks  on the evolution of
the GCMF}
Before discussing in detail the results obtained, we will make some
preliminary qualitative considerations on
the possible evolutionary paths of the GCMF
depending on the initial conditions chosen. In the following
discussion we will assume a log-normal initial GCMF.

 We start by noting that, for a given initial value of the cluster
mass, all evolutionary
processes, with the exception of 
 mass loss associated to stellar evolution which
depends only on 
the initial stellar mass function, are more efficient at smaller
distances from the  
Galactic center (see e.g. Vesperini 1997); on the other hand, at a
given galactocentric 
distance, disruption of clusters due to 
escape of stars through the tidal boundary is proportional to the
relaxation time and thus it is more
efficient for low-mass clusters while the efficiency of dynamical friction
is an increasing function of the mass of the cluster. This implies that,
while in any case inner regions are  those where
the effects of evolution are stronger, it is not obvious, {\it a
priori}, in what direction the difference between the initial and the
final  GCMF and the difference between the  GCMF of  inner and outer
clusters is to  be driven by
evolutionary processes; in fact, even though evolutionary
processes always tend to decrease the mass of individual globular
clusters, this obviously does not imply that the final mean value of
the GCMF will be smaller than the initial one. Since a number of
clusters will be disrupted before one Hubble time and will not be
part of the final system, depending on the balance between the
evolution of the masses of surviving clusters and the distribution of masses
of those disrupted,  the final mean value of the GCMF can be larger or
smaller than the initial one.

For the initial conditions chosen in Vesperini (1997), as a result of
disruption of inner low-mass clusters,  the difference was
always in the sense of inner clusters to be on the average more
massive than the outer ones, 
and the final mean value of the whole sample of
clusters was larger than the 
initial one but it is easily conceivable an initial GCMF
dominated by high-mass clusters whose evolution is  dominated by
the effects of dynamical friction and in which the final difference
between inner and outer clusters and the difference between the final and
the initial  mean value of the mass distribution is in the
opposite sense. 

As for the  dispersion of the GCMF, one can anticipate that
the effects of evolutionary processes is that of leading to a decrease
of this 
if the initial GCMF contains a significant fraction 
both of low-mass clusters, which are mainly affected by tidal disruption,
and of high-mass clusters
significantly affected by dynamical friction: in this case dynamical
friction and tidal disruption  deplete the tails of the initial GCMF
and thus make the dispersion of the sample of surviving
clusters smaller.
On the other hand the evolution of a
distribution with a very small initial dispersion will be
characterized by an increase of the dispersion, due to an
asymmetric diffusion driven by the different mass loss of clusters all
approximately with the same initial mass but located at different
distances from the Galactic center.

We  summarize in Table 1a the expected evolution of $\mm$ and $\sg$
under the effects of disruption by dynamical friction, disruption by
evaporation, and mass 
loss of individual clusters that do not suffer complete disruption, 
each one considered separately from
others. Depending on the relative efficiency of these three processes
the GCMF can evolve in four different directions as indicated in Table
1b. We can thus divide the space of initial parameters in four regions
according to the way $\mm$ and $\sg$ evolve and we note here that if
there is a common point  
among these four regions this will define the initial 
parameters of an ``equilibrium'' GCMF which will keep its initial
parameters unaltered after one Hubble time.

\subsection{Evolution of the GCMF}
The set of  initial values for  the dispersion
and mean value of  the GCMF, $\sigma_i$ and $\langle \log M
\rangle_i$, considered in our investigation is shown in figure 1a
and the corresponding  final values of
$\sigma$, $\sg_f$, and of $\mm$, $\mm_f$,  calculated at $t=15$ Gyr 
(hereafter 
by final value of any quantity we will mean
that calculated at this time)
are shown in figure 1b.  The parameters to describe the final GCMF are
estimated as described in sect.2.

Figures 2a and 2b show the contour plots of  $(\mm_f-\mm_i)$ and
$(\sg_f-\sg_i)$ in the plane $\mm_i-\sg_i$ from which it is clear the
dependence 
on the initial conditions of these quantities; figures 2c and 2d show
the contour 
plots of $\mm_f$ and $\sg_f$ in the same plane. In figure 3 we have
plotted only the curves corresponding to $\mm_f-\mm_i=0$ and $\sg_f-\sg_i=0$
which divide the plane of initial parameters in the four regions
described qualitatively in section 3.1 (see Table 1b). 

For low  values of $\sg_i$ the evolution is toward
larger final dispersions, while as $\sg_i$ increases the effects of
evolution drive the system toward values of the dispersion  smaller than the
initial ones; the transition occurs approximately at $\sg_i \simeq
0.65$. 
As for  $\mm$, for all the systems with $\mm_i>5.2$  disruption
by dynamical friction of high-mass clusters and  mass loss of
clusters without complete disruption are the dominant processes and
thus $\mm_f<\mm_i$; for $\mm_i<5.2$, $\mm_f$ can be smaller than
$\mm_i$ if the process of mass loss of clusters without complete
disruption dominates or larger than $\mm_i$ if disruption by complete
evaporation of low-mass clusters is the most important process.

The intersection of the
two curves, as anticipated in section 3.1, corresponds to the initial
(and final) parameters  of an equilibrium GCMF; this particular GCMF
(hereafter E-GCMF), as we will  show in detail below, has the very
interesting property of maintaining its initial shape and parameters
unchanged during the entire evolution even though a significant number
of clusters are disrupted because of evaporation or dynamical
friction.

Figures 4a-b show  the initial and the final GCMF with the corresponding
gaussian fit for two typical cases of evolution of the GCMF while
figure 4c shows the evolution of the E-GCMF.

It is important to note from this figure  that 
all the final GCMFs shown are still well described by a
gaussian distribution. This result, as pointed out in Vesperini
(1997), is far from being obvious given the large
number of clusters undergoing disruption during one Hubble time in
most of the cases considered. 

The qualitative scenario anticipated in section 3.1 is now clearly shown in
figure 4:
in GCMFs initially dominated by low-mass clusters  
the low-mass tail is significantly depleted and the final value of
$\mm$ is larger than the initial one (figure 4a); for initial GCMFs
dominated by high-mass clusters the evolution is in the opposite
direction as  the high-mass tail is that
more affected by evolutionary processes and the final value of $\mm$ is smaller
than the initial one (figure 4b).  Figure 4c shows the
E-GCMF: while, as it is evident from the figure,  the final
number of clusters in the system is significantly smaller than the
initial one (about 52 per cent of the clusters initially in the system
are disrupted) the shape and the parameters of the GCMF are
almost exactly 
preserved  after one Hubble time through the balance between
disruption of clusters and evolution of the masses of those which survive.
The initial dispersion and mean value of the GCMF  having this interesting
characteristic are $\sg\simeq 0.64$ and $\mm \simeq 4.93$.
The agreement with the values for the equilibrium GCMF obtained by
Vesperini (1997) ($\sg\simeq 0.5$ and $\mm=5.0$; see 
his GAU2 simulation) is quite remarkable given the
differences in the method adopted for the investigation in that and in
the present work and the fact that in Vesperini (1997) no systematic
investigation of different initial conditions were done to locate the
exact parameters of the equilibrium GCMF.
Figure 5  shows the  distribution of $\log M_i$ of those clusters in
the E-GCMF
which at $t=15$ Gyr have $\log M_f \simeq \langle \log M\rangle_f$ and
it clearly
illustrates that the equilibrium is preserved dynamically.

\subsection{Dependence of the GCMF on the galactocentric distance}
As we discussed in the introduction, investigation of
the possible differences between the properties of clusters located
closer to the Galactic center and the properties of those in the outer
regions of the 
Galaxy is of particular interest because it can provide important
clues on the actual role of evolutionary processes in determining the
present properties of globular clusters and on the reliability of the
distances of external galaxies estimated by using the turnover magnitude
of the GCLF of their globular cluster systems.

In our investigation we have classified as inner clusters all those
at a distance from the Galactic center, $R_g$,  smaller than 8 Kpc
and as outer clusters all those having $R_g>8$ Kpc.

First we focus our attention on the difference, $\Delta
\mm_{in-out}=\mmi-\mmo$,  between  the mean
value of the GCMF of inner clusters, $\mm_{inner}$, and that of the
GCMF of outer clusters , $\mm_{outer}$.

Figure 6 shows the plot of $\Delta \mm_{in-out}$ versus $\mm_i$.  
Depending on the initial
conditions, $\Delta \mm_{in-out}$ can be positive or negative:
 all systems with $\mm_i <
4.9-5.1$, the exact value depending on $\sg_i$, have, after one Hubble
time, $\mmo<\mmi$ while the opposite trend is
established by evolutionary processes  for systems initially dominated by
massive clusters.
The qualitative
explanation for the observed behaviour of $\Delta \mm_{in-out}$ is similar to
that discussed for the difference between the final and the initial
value of $ \mm$ for the whole sample of clusters and in fact, as shown
in figure 7, there is an evident correlation between $\Delta
\mm_{in-out}$ and $\mm_f-\mm_i$ : the reason for this
is clear  if one considers
that the properties of outer clusters are in general expected to
resemble those of the initial system while inner clusters are
likely to be those mainly responsible for the variation of the
properties of the cluster system since they are more affected by
evolutionary processes.

We point out that most of the systems considered evolve significantly,
undergoing a  strong depletion of their initial number of clusters, but in many
cases the difference  $ \Delta \mm_{in-out}$ induced by evolutionary
processes is not 
very large and it is easily conceivable that its observational
detection can be difficult. We will return to this point below in this
section and in section 5 when we will compare our results to
observational data for the Milky Way.

For most of the initial
conditions considered in this study, evolutionary processes tend to
make the dispersion of the  sample of inner clusters smaller than that
of outer clusters and this is consistent with the results of the
analysis of observational data for the Galaxy, M31 and M87 by Gnedin
(1997) and Kavelaars \& Hanes (1997). 
Only for systems initially characterized by a low
dispersion ($\sg_i < 0.5$)  the opposite trend is
established in the course of evolution as it is clear from figure
8 which  shows the plot of  $\Delta
\sg_{in-out}$ versus $\sg_i$. 
This result can be easily interpreted if, as already discussed in
section 3.1, one considers that 
for a GCMF with initial dispersion 
large enough, evolutionary processes, more efficient in the inner
regions, tend to deplete more efficiently the low-mass tail (by tidal
disruption) and 
the high-mass tail (by dynamical friction) of the initial distribution
thus causing it to become narrower as evolution goes on. 

While the above analysis based on the division of a cluster population
in two subpopulations of inner and outer clusters allows to easily
obtain a general  quantitative measure of the radial
variation of the properties of a GCMF, it is also important  to study
some cases more in detail and to address the issue of the radial
variation of the GCMF by making a finer division of clusters according
to their distance from the Galactic center and studying the properties
of GCMF in more than two radial bins. Thus we  
discuss now in some detail  to what  extent a  correlation 
between the galactocentric distance $R_g$ and the mean mass 
$\langle \log M\rangle_{R_g}$ and dispersion $\sg_{R_g}$
 of clusters located in spherical shells
with limits $R_g-\Delta R_g$ and $R_g+\Delta R_g$ can be produced by
evolutionary process. 
No significant correlation  between $R_g$ and $\langle
\log M\rangle_{R_g}$ is observed in many galaxies (see e.g. Harris et al. 1998,
Forbes, Brodie \& Hucra 1997, Forbes et al. 1996a,b), and this result is often
interpreted as an indication that evolutionary
processes do not play a relevant role in the evolution of globular
cluster systems. 

 We will show that a strong effect of evolutionary processes
does not necessarily imply the formation of  a strong radial gradient of
$\langle \log M \rangle_{R_g}$.

We focus our attention on three different initial log-normal GCMF among those
considered in section 3.2:\\
(a) $\langle \log M\rangle_i=4.6$, $\sigma_i=0.9$;\\
(b) $\langle \log M\rangle_i=5$, $\sigma_i=0.7$;\\
(c) $\langle \log M\rangle_i=5.8$, $\sigma_i=0.9$.\\
Initial conditions (a) and (c) have been chosen because they  are among those
more significantly affected by evolutionary processes since they contain  many
low-mass and many high-mass clusters respectively,  
 while (b), besides being a system undergoing a significant disruption
and decrease in the total number of clusters, is an initial condition
which could be similar to that of the 
Galactic globular cluster system (see sect. 5).
   As shown in fig.4 (panels
a and b) where the initial and the final GCMF  corresponding to the initial
conditions (a) and (c)  are plotted (the evolution of the system (b)
is not shown  in figure 4 but the initial conditions and the evolution are very
similar to those shown in fig. 4c for the E-GCMF), in all these three cases
evolutionary processes play a significant role leading to the disruption of a
significant number of globular clusters. Figure 9, in which we have
plotted   the initial and the final histogram of
$R_g$ for these three systems, clearly shows the depletion of clusters in the
inner regions of the Galaxy. 

Figure  10 shows  the plot of $\langle \log M\rangle_{R_g}$ and
$\sg_{R_g}$ versus $R_g$.  $\langle \log M\rangle_{R_g}$ and
$\sg_{R_g}$  have been calculated in five
different radial bins each one including clusters between $R_g-1.9$ Kpc and
$R_g+1.9$ Kpc with $R_g=2.9,6.7,10.5,14.3, 18.1$. It is evident that a
significant disruption of clusters is not necessarily followed by the
formation of a  strong  radial trend of
$\langle \log M\rangle_{R_g}$;  as we pointed out above, initial
conditions (a) and (c) are two rather extreme cases unlikely to be
relevant for real globular cluster systems and they have been chosen
because they are probably those able to give rise to the strongest
radial gradient of 
$\langle \log M\rangle_{R_g}$.  Initial condition (b), which is likely
to be a more realistic choice for the initial GCMF, does not have a
strong radial gradient of $\langle \log M\rangle_{R_g}$.
Moreover we point out that the detection of any
radial gradient of $\langle \log M
\rangle$ in globular cluster systems of external distant galaxies is likely to
be more difficult if  the sample of clusters observed in
distant galaxies does not include the low-mass tail of the GCMF.  This is clear
from figure 11 where we have plotted $\langle \log M\rangle_{R_g}$ versus
$R_g$ for the case (a) as in figure 10  but excluding the low-mass
tail of the 
GCMF beyond $1 \sigma$:
the  trend present when the complete population of clusters is
considered, 
disappears when the low-mass tail of the distribution is excluded. Finally we
note that, since we  have considered only circular orbits, we expect any
radial gradient in our theoretical sample  to be stronger than the
corresponding one derived by observational data where the current values of the
galactocentric distances do not necessarily provide an exact indication of the 
galactic tidal field affecting  clusters; this, of course, does not
mean that no gradient at all  is expected when non-circular orbits are
considered and in fact Murali \& Weinberg (1997) and Baumgardt (1998),
who include non-circular orbits in their analysis, find that
evolutionary processes give 
rise to a radial trend in the properties of the mass distribution with
clusters on more eccentric orbits being in general those more easily
disrupted. The  situation is even more 
unfavourable  to the detection of a  radial gradient when, as in the case of
globular clusters in external galaxies, observational data provide projected
distances from the galactic center.

In conclusion while in principle evolutionary processes are expected to induce
a radial gradient in the mean mass of globular clusters, in practice
the extent of this depends significantly on initial conditions and in
some cases, even though evolutionary processes are very efficient and a
significant number of clusters are disrupted, the 
radial gradient of $\langle \log M\rangle $ is very weak. 

\subsection{Time evolution of the GCMF}
While in the previous sections we have focussed our attention on the final
properties of the GCMF and investigated their dependence on the
initial conditions, in this section we will study the  time
evolution of the parameters of  the GCMF by showing and discussing some
typical ``paths'' followed by the properties of the GCMF from $t=0$ to
$t=15$ Gyr.

The rate of evolution of the GCMF properties is determined by the
rate of  disruption of
clusters and the rate of change of the masses of those which survive.
Mass loss by stellar evolution depends on time and 
it is significant only for $t <1 $Gyr (see e.g. Vesperini
\& Heggie 1997) when more massive stars  evolve out of the
main sequence; as for disruption by evaporation and by
dynamical friction, 
since, for a given mass, inner clusters are those expected to be more
affected by these evolutionary processes,  the rate of evolution of the
GCMF is to depend on the number of clusters closer to the Galactic
center; as time goes on and the ``weaker'' clusters closer to the
Galactic center are disrupted by evaporation or dynamical friction, the
number of clusters  liable to the effects of these processes on
relatively short time scales decreases
and eventually the rate of change in GCMF parameters  slows
down.

Figures 12a-c , where we
have plotted 
the time evolution of $\mm$,  of $\sg$ and of the ratio of the total
number of clusters in the system at time $t$ to the total initial
number of clusters clearly show
this effect: after a phase
of significant change in $\mm$ and $\sg$, as mass loss by stellar
evolution ceases to be important and the number of clusters in the
inner regions of the Galaxy has decreased, the evolution of
the parameters of the GCMF  slows down significantly.

It is important to note the difference between the equilibrium reached
due to a more ``favourable'' spatial distribution and the more
interesting E-GCMF
discussed in section 3.2  which remains unchanged since the beginning
in  dynamical equilibrium independent on the underlying spatial distribution.

Figure 13 further clarifies this point by  clearly showing the existence
of a particular GCMF able to stay in equilibrium for one Hubble
time. The trajectories shown in figure 13 are determined in the
following way: for a given initial condition an arrow joins the
initial conditions in the plane $\sg$-$\mm$ to the corresponding
values after 15 Gyr, these final values of $\sg$ and $\mm$ are then used
as initial conditions for the following step in which, again, an arrow
joins these new initial conditions to the corresponding final values
and so on. It is important to realize that the trajectory obtained in
this way is not a real trajectory  in the plane $\sg$-$\mm$ since each
time a new step is done, the
starting point of the new step is the final state of the previous one
but the underlying changes in the spatial distribution of clusters in
the Galaxy and in the
stellar mass function of individual globular clusters, which cause
 the slowing down of the
evolution, are eliminated; thus the system restarts with the full spatial
distribution, each cluster loses mass by stellar evolution 
according to the complete initial stellar mass function and a new
impulse is given to the motion in 
the $\sg$-$\mm$. In this sense the E-GCMF is an attractor in the plane
$\sg$-$\mm$ since it is the 
only GCMF which, taken as an initial condition, does not need any
particular spatial distribution and the decrease of mass loss
associated to stellar evolution to slow the rate at which its
parameters evolve; as discussed above, these preserve their initial
values unaltered for a Hubble time  due to the balance between
disruption of clusters and evolution of the masses of those which
survive.

\subsection{Fraction of surviving clusters}
One interesting issue concerning the Galactic globular cluster system
is the relationship between their present number,
 their present total mass and the corresponding
initial values of these quantities. As expected, the fraction of
surviving clusters and the ratio of 
the total present mass of clusters to the initial one depend
significantly  on the
initial conditions. 
Figures 14a and 14b show the contour plot of the ratio of the total
number of clusters surviving after 15 Gyr to the total initial number
of clusters, $F_N$, and of the ratio of the total mass of all surviving
clusters after 15 Gyr to the total initial mass of all clusters,
$F_M$, in the plane $\sg_i$-$\mm_i$. 

For a given value of $\sg_i$, $F_N$ ($F_M$ has a similar
 behaviour) has a maximum
for the value of $\mm_i$, $\mm_{max}$,
corresponding to  the most ``robust'' initial
GCMF which is that having the minimum number of clusters undergoing
disruption by evaporation (low-mass clusters) or by dynamical friction
(high-mass clusters); the increasing number of low-mass clusters
(easily disrupted by evaporation), for $\mm_i< \mm_{max}$ and of
high-mass clusters (undergoing disruption by dynamical friction)
for  $\mm_i> \mm_{max}$ explains the observed decrease of $F_N$.

Figure 14 gives an indication on the total number of clusters
disrupted in one Hubble time but it is also interesting to estimate
the current disruption rate, $F_D$, defined as 
the fraction of the number of clusters at $t=15 $ Gyr undergoing
disruption in the next 1 Gyr
\be
F_D={N_{gc}(15)-N_{gc}(16) \over N_{gc}(15)}
\ee
where $N_{gc}(t)$ indicates the total number of clusters at time $t$
(in Gyr). Hut \&
Djorgovski (1992) by an analysis of the distribution of the
half-mass relaxation times of Galactic clusters have estimated $F_D
\simeq 0.038 \pm 0.02$. Figure 15 shows the contour plot of $F_D$ in
the plane 
$\sg_i$-$\mm_i$. It is interesting to note that for the values of $\sg_i$
and of $\mm_i$ which are likely to be relevant for the Galactic
globular cluster system ($\mm \simeq 5$ and $\sigma \simeq 0.7$)
 the values of
$F_D$ we obtain are very close to that obtained by Hut \& Djorgovski (1992).
\subsection{Spatial distribution}
In section 3.4 we have already made some comments on the evolution of
the spatial distribution of clusters in the Galaxy, pointing out that
the larger efficiency of evolutionary processes in the inner regions
of the Galaxy tends to flatten the profile of number density  of
clusters close to the galactic center. In the outer regions the
fraction of disrupted clusters is in most cases negligible and the
initial power-law profile with index equal to -3.5 is preserved. In
order to provide a quantitative measure of the flattening in the
density profile acquired during the evolution as a function of the
initial conditions we have calculated the core radius $R_c$ by fitting
the final number density profile for all the
cases investigated with the following function
 (see, e.g. Djorgovski \& Meylan 1994)
\be
n(R_g)=A\left(1+R_g/R_c\right)^{-3.5} \label{ndd}
\ee
We note that for  the initial conditions $n(R_g) \propto R_g^{-3.5}$
between $R_g=1$ Kpc and $R_g=20$ Kpc  a fit of the initial number
density by eq. (\ref{ndd}) results in  a very small core radius
($R_c\simeq 0.1$ Kpc).

The contour plot of $R_c$
in the plane $\sg_i$-$\mm_i$ is shown in fig.16.
In all the  cases investigated the spatial distribution tend to
flatten in the inner 
regions of the Galaxy and the effect is stronger (larger values of
$R_c$) for initial 
conditions containing many low-mass clusters (efficiently disrupted by
evaporation) or many high-mass clusters (efficiently disrupted by
dynamical friction). 

\subsection{Inclusion of disk shocking}
All the simulations discussed in the previous section do not include
the effects of disk shocking. In the light of the results obtained by
Vesperini \& Heggie (1997) concerning the difference between the
time evolution of the total mass with and without disk shocking, it is
obvious to expect some 
quantitative differences in the final results  once the effects of disk
shocking are included but all the general conclusions concerning the
trends and the dependence of the final results on the initial
conditions, the existence of an equilibrium GCMF, the definition of
different regions in the plane of initial conditions are 
unaltered.

Since,  as
discussed in Vesperini \& Heggie 
(1997), the analytical expression providing the time evolution of the
total mass of a cluster as a function of its galactocentric distance
and its initial mass is more approximate and is valid for a smaller
range of initial parameters than the corresponding one without the
effects of disk shocking, we have adopted this formula only for
clusters with $R_g<8$ Kpc assuming the effects of disk shocking at
larger galactocentric distances to be negligible.

Figures 17a-d show the plots of the final values of $\mm$, $\sg$,
$F_N$ and $F_D$ from  the runs not including disk shocking
versus the corresponding
values ($\mm_{f,ds}$,$\sg_{f,ds}$,$F_{N,ds}$, $F_{D,ds}$) 
 obtained from the simulations  including disk shocking.
 As expected in all cases the number of surviving clusters is smaller
when disk shocking is included but $F_N-F_{N,ds}$ is never larger than
$0.2F_N$ and in most cases is smaller than $0.1 F_N$. As for 
 $\mm$ and $\sg$ we obtain $-0.015<(\mm_{f,ds}-\mm_f)/\mm_f<0.013$ and
$-0.12<(\sg_{f,ds}-\sg_f)/\sg_f<0.07$.

For what concerns the parameters of the equilibrium GCMF we obtain
$\mm_i\simeq 5.03$, $\sg \simeq 0.66$ quite close to the corresponding values
obtained without disk shocking ($\mm_i\simeq 4.93$, $\sg \simeq
0.64$).

\section{Different initial GCMF}
A complete theory of the process of globular cluster formation
is still lacking and the only support for a log-normal  initial GCMF
comes from the observations of the present GCMF. While assuming such a
functional form 
also for the initial GCMF is a reasonable guess it is interesting to
consider other possibilities.

In some works it has been suggested that the initial GCMF could 
be a power-law in $M$ (see e.g. Harris \& Pudritz (1994), Elmegreen
\& Efremov (1997) 
and also Elemegreen \& Efremov (1997) for   references to
past observational and theoretical analysis on this issue) and
observations of the luminosity function of young clusters in
interacting galaxies (see e.g. Whitmore \& Schweizer (1995) for NGC
4038/39) show a GCLF which is a power law in L (but see
Fritze-v.Alvensleben (1998), for a recent interesting analysis
determining the intrinsic mass distribution of young clusters in NGC
4038/39 and 
showing this to be a log-normal distribution; see also Meurer (1995)). 
Thus we
have considered this possibility and investigated the evolution of
systems with an initial GCMF given by
\be
f(M)= AM^{-\alpha} \hbox{ for } M_{low}<M<M_{up}
\ee
with
$\alpha=2,1.7,1.5$, $M_{up}(M_{\odot})=10^{6.5}$ and
$M_{low}(M_{\odot})=10^3,~10^{3.5},~10^4,~10^{4.5},$ $~10^{5}$, $~10^{5.5}$.
In agreement with the results of previous analysis (see e.g. Okazaki
\& Tosa 1995, Vesperini 1994, 1997, Baumgardt 1998) we obtain
that 
in all these cases evolutionary processes tend
to modify the initial power-law  GCMF in a bell-shaped GCMF when
binned in $\log M$. 
In figure 18 the initial and the final GCMF for the case
($M_{low}=10^4 M_{\odot},~\alpha=2$) are  shown: it is evident that
evolutionary processes are very efficient in modifying the initial
GCMF into a log-normal distribution. As $M_{low}$ increases the
evolution of the mass of individual surviving clusters occurs on
longer timescales and thus the process of asymmetric diffusion (see
section 3.1) which is responsible for the formation of the low-mass tail
in the final GCMF is slower; consequently, even though the tendency of
the GCMF to evolve toward a bell-shaped distribution in $\log M$
persists, the deviations of the GCMF at $t=15$ Gyr from a log-normal
distribution increases for larger values of $M_{low}$.

Figures 19a-b show the plots of $\mm$ and $\sg$ of the final GCMF
(defined as described in sect. 2 also for the cases in which the final
GCMF deviates significantly from a log-normal distribution)  versus
$\log M_{low}$ for the values of $\alpha$ considered. It
is interesting to note the existence of two regimes: for small values
of $M_{low}$ ($M_{low}<10^{4.5}M_{\odot}$), the final value of $\mm$ and $\sg$
do not depend on  $M_{low}$ and they are determined  by
evolutionary processes and by the value of $\alpha$;
 for $M_{low}>10^{4.5}M_{\odot}$, $\mm$ and
$\sg$ are essentially determined by
the value of $M_{low}$ and  $\mm$ keeps memory of the
initial value of $M_{low}$ (in fact it is approximately equal to
$\log M_{low}$).

\section{Comparison with observational data }
As we discussed in the introduction 
 a detailed comparison of our theoretical results with observational
data is beyond the scope of this work both because  of some
 limitations in our analysis and because of the uncertainties in the
 initial conditions to adopt for the Galactic globular cluster system.
Nevertheless, while keeping these caveats in mind, it is interesting
 to see to what extent the predictions of our analysis are in general
 consistent with some of the observed properties of the Galactic
 globular cluster system.

Real initial conditions of the Galactic globular cluster system are
unknown but a reasonable assumption is that of looking at the
properties of  clusters in the outer regions of the Galaxy, where
evolutionary processes act on longer timescales,  and adopting these as
initial conditions for the 
entire cluster system (see also Fall \& Malkan 1978 for an interesting
study aimed to obtain information on the initial distribution of
globular  cluster core properties from the current observational data).
From the estimate of the mean magnitude
and dispersion of the current GCLF for galactic halo  clusters with
$R_g>8 \hbox{ Kpc}$ obtained by Gnedin (1997) (see his table 1), assuming
$M/L_V=2$, the following values are obtained: $\mm_i=5.0$,
$\sg_i=0.7 $. Assuming an initial log-normal GCMF with these
parameters we have calculated some quantities that can be compared
with observational data.

The results of this calculation, summarized in Table 2, are in general
in good agreement with the observational data.

In agreement with the  analysis  by Kavelaars \& Hanes
(1997), Gnedin (1997) and Ostriker \& Gnedin (1997), we find that the
dispersion of inner clusters
is smaller than that of outer clusters; as for  the
difference between $\mm$ of inner and outer 
we find $\mm_{inner}>\mm_{outer}$ in agreement with the conclusion of 
Gnedin (1997) and Ostriker \& Gnedin (1997). 

According to our results the current number of clusters would be about
half ($F_N=0.54$ if disk shocking is not considered and $F_N=0.48$
taking into account the effects of disk shocking) of the total initial
number of clusters and the total current mass would be about 40 per
cent of the total initial mass. This implies that  
the total initial number of clusters in our Galaxy would be about 300 and 
their total initial mass  about 
$9\times 10^7 M_{\odot}$ (where we have adopted $M/L_V=2$ and the
values of the total visual magnitude of Galactic clusters tabulated in
Djorgovski 1993) from which it follows that the contribution to the halo mass
from disrupted clusters and stars escaped from survived clusters 
would be about $5.5\times 10^7 M_{\odot}$ a value very far from being
able to account for  the entire halo mass.

Figure 20 shows the histogram of the ratio of the current to
initial masses of surviving clusters (the effects of disk shocking are
considered) providing us with an interesting information on the
relationship between the initial and the final state of surviving
clusters. 
The bin $0.9<M_f/M_i<1$ is empty since all clusters lose about 18 per
cent of their initial mass due to the mass loss associated to the
stellar evolution; this means that clusters from which no star escapes
have $M_f/M_i \simeq 0.82$ and  they are in the bin
$0.8<M_f/M_i<0.9$. It is interesting to note that, even though the
masses of many surviving clusters are quite close to their initial
values, there are several clusters which have lost  a significant
fraction of their initial mass; for  about 40 per cent of the
surviving clusters the current mass is less than half the initial
mass. As discussed in Vesperini \& Heggie (1997) the stellar IMF of
these clusters evolves significantly and a significant fraction of
their current mass is expected to be in white dwarfs and thus they are
object of great interest; unfortunately most of them are likely to be
located in the inner regions of the Galaxy and they are challenging
targets for observations.

We conclude this section discussing the issue of the distribution  of
the disruption timescale $t_d(t)$, defined as the time necessary for
the complete disruption of a cluster at a time $t$ after its
formation.

In a recent investigation Gnedin \& Ostriker (1997) have addressed
this point and have estimated the destruction rate for 119 Galactic
globular clusters taking into account the current observed properties
of the clusters considered.

As discussed in detail in Gnedin \& Ostriker (1997), assuming that all
clusters were formed at the same time $t_0\simeq 0$ the number of
clusters surviving at time $t$ is equal to the number of clusters for
which the initial value of $t_d$ was larger than $t$ and the current
distribution of disruption timescales is connected to the initial one
by the simple relation $f(t_d,t_H)=f_i(t_d+t_H)$, where $f_i$ denotes
the initial distribution of the disruption timescales.

In figure 21a we show the histogram of $\log(t_d(t_H)/t_H)$ which we have
obtained for the sample of clusters surviving at $t_H$ from an initial
population with a GCMF equal to that adopted above in this section for
the comparison of our data with observational data. The effects of
stellar evolution, disk shocking, two-body relaxation and dynamical
friction have been considered and, for ease of later comparison with
the results of the analysis of Gnedin \& Ostriker, the Hubble time has
been taken equal to $t_H=10$ Gyr.

If, as suggested by Gnedin \& Ostriker (1997), we make the hypothesis that
the initial distribution of $t_d$ is a power-law, $f_i(t_d) \sim
t_d^{-q}$, from the median value of $t_d$ in our sample we estimate
$q\simeq 1.6$ which falls in the range of the values obtained by
Gnedin \& Ostriker and as shown in figure 21a (dashed line) this
distribution fits fairly well the data. On the other hand, since in
our analysis we know the real initial conditions we can calculate the
initial distribution of $t_d$ and verify the hyopthesis made. In
figure 21b the distribution of $t_d(t=0)$ is shown and it is evident
that the initial distribution is not a power-law, but it is well
fitted by a log-normal distribution; as expected, adopting this
functional form for $f_i(t_d)$ a much better fit is obtained for
$f(t_d,t_H)$ (solid line in figure 21a).

We conclude emphasizing that much caution is needed in drawing any
conclusion on the 
initial number and initial properties of clusters on the basis of the
current distribution of disruption times; as shown in fig.22, in fact,
the differences in $f(t_d,t_H)$ derived from three very different initial
populations can be quite small and even a small uncertainty in the
estimates of the parameters of the current distribution of timescales,
or in the timescales themselves, can lead to a large error in the
estimate of the parameters of the initial population of clusters.

\section{Summary and conclusions}
In this work we have investigated the evolution of the mass function of a
globular cluster system located in a model for the Milky Way. The effects of
stellar evolution, two-body relaxation, disk shocking, dynamical friction
and  the presence of the tidal field of the Galaxy have been taken into
account in the evolution of the mass of individual globular
clusters in the system which is calculated on the basis of the results of the
$N$-body simulations carried out by Vesperini \& Heggie (1997).

A log-normal and a power-law initial GCMF have
been considered.
The main effort has been devoted to the investigation of systems starting with
an initial log-normal GCMF spanning a wide range of values of the
mean value and  the dispersion of the initial distribution. 
The gaussian shape has been shown
to be preserved very well during the entire evolution until $t=15 $ Gyr  while
for systems in which  the initial GCMF is a power-law  a  
bell-shaped GCMF resembling a Gaussian in $\log M$ tends to be
established in the  course of evolution.

Depending on the initial GCMF parameters, the mean value of the GCMF, $\langle
\log M \rangle $, can increase or decrease during the evolution according to
whether  the dominant process  is that of disruption of low-mass clusters by
evaporation of stars through  the tidal boundary (for initial GCMF dominated by
low-mass clusters) or that of disruption by dynamical friction of high-mass
clusters (for initial GCMF dominated by high-mass clusters). The regions in the
space of initial parameters   $\langle \log M \rangle_i $-$\sigma_i$
corresponding to these two different regimes  as well as the corresponding
regions for the evolution of the dispersion $\sigma$ has been
shown.

The differences between the final values of $\langle \log M \rangle$, $\Delta
\mm_{in-out}$,  and
$\sigma$, $\Delta \sigma_{in-out}$, of inner ($R_g<8 $ Kpc) and outer clusters
($R_g>8$ Kpc) have been investigated. 
Depending on the dominant evolutionary process ( 
disruption of low-mass clusters or dynamical friction) 
$\Delta \mm_{in-out}$ can be larger or
smaller than zero. As for  $\Delta \sigma_{in-out}$, in most cases considered
evolutionary processes tend to make the dispersion of inner clusters smaller
than that of outer clusters. The formation of a gradient of $\mm$ and
$\sigma$ with the galactocentric distance due to evolutionary
processes has been investigated. The direction of the gradient of
$\mm$ depends on the initial GCMF: an increasing $\mm$ as $R_g$
increases is common for systems initially containing many high-mass
clusters while the opposite trend is typical of systems initially dominated by
low-mass clusters. It has been shown that a significant effect of 
evolutionary processes does not necessarily imply the formation of a
strong radial trend of
$\langle \log M\rangle$. 

For most 
initial conditions considered, and in particular for those likely to
be relevant for real systems, evolutionary processes give rise to a
trend for $\sigma$ to decrease at smaller galactocentric distances.

The existence of a particular GCMF able to stay in dynamical equilibrium
keeping its initial shape and parameters unaltered during the entire
evolution by means  a subtle balance of disruption of clusters and evolution of
the masses of those surviving, first suggested by Vesperini (1997), has been
confirmed. 

The initial number density distribution of clusters in the Galaxy has
been taken 
proportional to $R_g^{-3.5}$ and it has been shown that evolutionary processes
tend to flatten this distribution close to the Galactic center. The extent of
the flattening depends on the initial conditions and it has been estimated
quantitatively by calculating the final core radius, $R_c$, of the distribution
of survived clusters in the Galaxy. The range spanned by $R_c$ for the  initial
conditions considered in our work is $0.4-2$ Kpc.

The fraction of the total initial number of clusters surviving after one Hubble
time, $F_N$,  the fraction  of the total initial mass of all the clusters in
the system, $F_M$,  and the current cluster disruption rate (defined as the
fraction of the number of clusters at $t=15 $  Gyr  undergoing disruption
within the next 1 Gyr), have been calculated and their dependence on
the initial 
conditions investigated.

The exact comparison of our results with observational
data is beyond the scope of our work both because of some simplifying
assumptions 
we have made and because of the current lack of a precise knowledge of the
initial properties of the Galactic globular cluster system; nevertheless
assuming the current properties of outer clusters to be similar to the initial
ones of the entire system we have calculated the values predicted
from our analysis for some of the observed properties of the Galactic globular
cluster system and we have
found them to be in general in good agreement with the  observational
values. 

As for the fraction of the total initial number of cluster surviving at
the current 
epoch, $F_N$, and the fraction of the total initial mass of all the globular
clusters, $F_M$, the values predicted for the Galactic system are $F_N \simeq
0.54$ (0.48 if the effects of disk shocking are included) and $F_M\simeq 0.41$
(0.38 with disk shocking). These values imply that the initial
population of Galactic globular clusters would consist of about 300
clusters with a total mass of about $9\times 10^7 M_{\odot}$ and
that the contribution to the halo mass from disrupted clusters and
stars escaped from survived clusters would
be about $5.5\times 10^7 M_{\odot}$.
The distribution of clusters disruption times has been calculated and
shown to be similar to the distribution of disruption times for
a sample of 119 Galactic clusters obtained by Gnedin \& Ostriker
(1997); we have shown that very different initial distributions of
disruption timescales can lead to very similar final distributions and
thus much caution is necessary in drawing any conclusion on the
initial population of clusters from the present distribution of
disruption timescales.

\section*{Acknowledgments}
I wish to thank Giuseppe Bertin and Rainer Spurzem for many useful
comments on this paper and D.C.Heggie and M.D.Weinberg  for many useful
and enlightening discussions.
\section*{References}
Abraham R.G., van den Bergh S., 1995, ApJ, 438, 218\\
Aguilar L. , Hut P.  Ostriker J.P. 1988, ApJ, 335, 720\\
Bahcall J.N., 1984, ApJ, 287, 926\\
Baumgardt H., 1998, A\&A, 330, 480\\
Bellazzini M., Vesperini E., Ferraro F.R., Fusi Pecci F., 1996, MNRAS,
279,337\\ 
Binney J., Tremaine S., 1987, Galactic Dynamics, Princeton University
Press, Princeton, New Jersey\\ 
Caputo F., Castellani V., 1984, MNRAS, 207,185\\
Chernoff D.F., Kochanek C.S., Shapiro S.L., 1986, ApJ, 309, 183\\
Chernoff D.F. \& Shapiro S.L., 1987, ApJ, 322, 113 \\
Chernoff D.F., Djorgovski S.G., 1989, ApJ, 339, 904\\
Chernoff D.F., Weinberg M.D., 1990, ApJ, 351, 121\\
Crampton D., Cowley A.P., Schade D., Chayer P., 1985, ApJ, 288, 494 \\
Djorgovski S.G., 1993 in Structure and Dynamics of Globular Clusters,
Djorgovski S.G., Meylan G. Eds., Astron. Soc. of the Pac. Conf. Series
vol.50, p. 373\\
Djorgovski S.G., Meylan G., 1994, AJ, 108, 1292\\
Elmegreen, B. G., Efremov, Y., 1997, 480, 235\\
Fall S.M., Malkan M.A., 1978, MNRAS, 185, 899\\
Fall S.M., Rees M.J., 1985, ApJ, 298, 18\\
Forbes D.A.,Franx M., Illingworth G.D., Carollo C.M., 1996a, ApJ, 467,
126\\
Forbes D.A., Brodie J.P., Hucra J., 1996b, AJ, 112, 2448\\
Forbes D.A., Brodie J.P., Hucra J., 1997, AJ, 113, 887\\
Fritze-v.Alvensleben U., 1998,preprint, astro-ph 9803139\\
Gnedin O.Y., 1997, ApJ, 487, 663\\
Gnedin O.Y., Ostriker J.P., 1997, ApJ, 474, 223\\
Harris W.E., 1991, ARAA, 29, 543\\
Harris W.E., Pudritz R.E., 1994, ApJ, 429, 177\\
Harris W.E., Harris G.L.H., McLaughlin D.E., 1998, preprint, astro-ph 9801214\\
Hut P., Djorgovski S.G., 1992, Nature, 359, 806\\
Jacoby G.H., et al. 1992, PASP, 104, 599\\
Kavelaars J.J., Hanes, D.A., 1997, MNRAS, 285, L31\\
Kissler-Patig, M., 1997, A\&A, 319,83\\
McLaughlin D.E., 1994, PASP, 106, 47\\
Meurer G.R., 1995, Nature, 375, 742\\
Meylan G., Heggie D.C., 1997, A\&A Rev., 8, 1\\
Murali C, Weinberg M.D., 1997, MNRAS, 291, 717\\
Okazaki, T., Tosa, M., 1995, MNRAS, 274, 48\\
Ostriker J.P., Gnedin O.Y., 1997, ApJ, 487, 667\\
Secker J, 1992, AJ, 104, 1472\\
van den Bergh S., 1995, AJ, 110, 1171\\
Vesperini E., 1994, Ph.D. Thesis, Scuola Normale Superiore, Pisa\\
Vesperini E., 1997, MNRAS, 287, 915\\
Vesperini E., Heggie D.C., 1997, MNRAS, 289, 898\\
Vietri M., Pesce E., 1995, ApJ, 442, 618\\
Whitmore B.C., Schweizer F., 1995, AJ, 109, 960\\
Wielen R., 1988, in The Harlow-Shapley Symposium on Globular Cluster
Systems in Galaxies, ed. J.E. Grindlay \& A.G.D. Philip, IAU Symp. 126
(Dordrecht, Reidel) p.393\\
Zinn R., 1985, ApJ, 293, 424
\newpage
\clearpage
\begin{table}
\begin{tabular}{|cccc|}
\multicolumn{4}{c}{\bf Table 1a}\\
\multicolumn{4}{c}{Effects of evolutionary processes}\\
\multicolumn{4}{c}{ on the parameters
of the mass function of a globular cluster system}\\
\hline
$\#$&process&$\sigma$&$<\log M>$\\
\hline
$I$&disruption by evaporation&$\downarrow $&$\uparrow$\\
$II$&disruption by dynamical friction&$\downarrow $&$\downarrow$\\
$III$&mass loss of individual surviving clusters&$\uparrow $&$\downarrow$\\
\hline
\multicolumn{4}{l}{A log-normal GCMF with dispersion $\sigma$ and mean
value $<\log M>$ is assumed.}\\
\multicolumn{4}{l}{The third and the fourth column indicate the
effect on $\sigma$ and $<\log M>$ respectively}\\
\multicolumn{4}{l}{if the corresponding
process indicated in the second column were the only one}\\
\multicolumn{4}{l}{determining the evolution of the GCMF.}\\
\end{tabular}
\end{table}
\begin{table}
\begin{tabular}{|ccc|}
\multicolumn{3}{c}{\bf Table 1b}\\
\multicolumn{3}{c}{Evolution of the parameters of the GCMF}\\
\hline
$<\log M>_f-<\log M>_i$&$\sigma_f-\sigma_i$&balance of evolutionary effects\\
\hline
$<0$&$<0$&$II_M+III_M>I_M$; $I_{\sigma}+II_{\sigma}>III_{\sigma}$\\
$<0$&$>0$&$II_M+III_M>I_M$; $I_{\sigma}+II_{\sigma}<III_{\sigma}$\\
$>0$&$<0$&$II_M+III_M<I_M$; $I_{\sigma}+II_{\sigma}>III_{\sigma}$\\
$>0$&$>0$&$II_M+III_M<I_M$; $I_{\sigma}+II_{\sigma}<III_{\sigma}$\\
\hline
\multicolumn{3}{l}{$I,II, III$ indicate the effect on the dispersion
(subscript $\sigma$)}\\
\multicolumn{3}{l}{ and on the mean value (subscript $<\log M>$) of
the  GCMF }\\ 
\multicolumn{3}{l}{ of the three evolutionary processes as indicated
in Table 1a}\\ 
\end{tabular}
\end{table}
\newpage
\clearpage
\begin{table}
\begin{tabular}{|cccc|}
\multicolumn{4}{c}{\bf Table 2}\\
\multicolumn{4}{c}{Comparison with observational data}\\
\hline
&Observ.&Theor$_{no-ds}$& Theor$_{ds}$\\
\hline
$\mm$&$ 5.10 \pm 0.06$&4.99&5.02\\
$\sigma$&$0.56\pm 0.04$&0.65&0.67\\
$\Delta \mm_{in-out}$&$ 0.16 \pm 0.09$&0.07&0.12\\
$\Delta \sg_{in-out}$&$- 0.09\pm 0.06$&-0.07&-0.06\\
$F_D$&$ 0.038\pm 0.02$&0.036&0.036\\
$F_N$&$-$&0.54&0.48\\
$F_M$&$-$&0.41&0.38\\
\hline
\multicolumn{4}{l}{Observational values of $\mm$, $\sg$ are from
Gnedin (1997) (assuming $M/L_V=2$),}\\
\multicolumn{4}{l}{ $\Delta
\mm_{in-out}$ and $\Delta \sg_{in-out}$ 
are taken from Ostriker \& Gnedin (1997).}\\
\multicolumn{4}{l}{The observational value of $F_D$ is taken from
Hut \& Djorgovski (1992).}\\
\multicolumn{4}{l}{Theor$_{no-ds}$ indicates the theoretical values
estimated without the}\\
\multicolumn{4}{l}{ effects of disk shocking.}\\
\multicolumn{4}{l}{Theor$_{ds}$ indicates the theoretical values
estimated with the}\\
\multicolumn{4}{l}{ effects of disk shocking.}\\
\end{tabular}
\end{table}
\newpage
\clearpage
\section*{Figure captions}
Figure 1  (a) Dispersion $\sigma_i$ and mean value of
$\log M$, $\langle \log M\rangle_i$ , of the initial log-normal GCMFs
investigated  
in this paper; (b) final (at $t=15 $ Gyr) values of $\sigma$,
$\sigma_f$, and of
$\langle \log M\rangle$, $\langle \log M\rangle_f$, from  the set of initial
conditions shown in (a).\\
Figure 2 Contour plots in the plane $\sigma_i-\langle \log M\rangle_i$ of 
$\langle \log M\rangle_f-\langle \log M\rangle_i $ (a), $\sigma_f-\sigma_i$
(b), $\langle \log M\rangle_f$ (c), $\sigma_f$ (d).\\
Figure 3  Initial values of $\sigma_i$ and $\langle
\log M\rangle_i$ for which $\sigma_f=\sigma_i$ (dashed line) and initial  
values of $\sigma_i$ and $\langle \log M\rangle_i$ for which $\langle \log
M\rangle_f=\langle \log M\rangle_i$ (solid line).
 These two lines divide the space of
initial conditions in four regions, each one characterized by a different
evolution of the parameters of the GCMF (see discussion in the text). The
point of intersection of the two lines  corresponds to an "equilibrium" initial
GCMF whose parameters do not evolve even though a significant number
of clusters  
are disrupted in a Hubble time (see text).\\
Figure 4  Initial (dashed lines) and final (dots and solid  lines) GCMF for
(a) $\sigma_i=0.9$, $\langle \log M\rangle_i=4.6$,
(b) $\sigma_i=0.9$, $\langle \log M\rangle_i=5.8$, (c) $\sigma_i=0.64$,$\langle
\log M\rangle_i=4.93$. Dots show the real GCMF obtained at $t=15 $ Gyr
while solid 
lines show the gaussian distributions with mean value and dipersion estimated
from the sample of surviving clusters as discussed in sect.2.\\
Figure 5 Distribution of the initial values of the mass of those clusters which
have  $\log M$ at $t=15$ Gyr approximately equal to $\langle \log
M\rangle_f $ (in the range $\mm_f\pm 0.03$). The 
initial GCMF is a log-normal distribution with $\sigma=0.64$ and
$\mm=4.93$ (E-GCMF see text).\\
Figure 6  Difference, $\Delta \mm_{in-out}$ between the final value
of $\langle \log 
M\rangle$ of inner clusters ($R_g<8$ Kpc) and that of outer clusters
($R_g>8$ Kpc) versus the initial mean value of the GCMF.\\ 
Figure 7  $\Delta \mm_{in-out}$ versus the difference between the
final and the initial value of $\langle \log M\rangle$ for all clusters.\\ 
Figure 8  $\Delta \sigma_{in-out}$ versus the initial dispersion of the GCMF
$\sigma_i$.\\
Figure 9 Initial (solid line) and final histogram of $R_g$ for three different
initial GCMF: $\sigma_i=0.9$, $\mm_i=4.6$ (dotted line);
$\sigma_i=0.7$, $\mm_i=5$ (short-dashed line); $\sigma_i=0.9$,
$\mm_i=5.8$ (long-dashed line).\\ 
Figure 10 (a) $\mm$ at $t=15$ Gyr for clusters in five radial bins versus
galactocentric distance of the bin for the following initial
conditions $\sigma_i=0.9$, $\mm_i=4.6$ (filled dots);  $\sigma_i=0.7$, $\mm_i=5$
(open circles); $\sigma_i=0.9$, $\mm_i=5.8$ (triangles).
(b)   $\sigma$ at $t=15$ Gyr for clusters in five radial bins versus
galactocentric distance of the bin. (symbols as in (a)).\\
Figure 11 Same as figure 10 (a) for the initial condition
$\sigma_i=0.9$, $\mm_i=4.6$ but excluding the low-mass tail of the
GCMF at $t=15$ Gyr. (the values for the entire sample already shown in
figure 10 (a) are plotted as filled dots for ease of comparison).\\
Figure 12 Time evolution of $\langle \log M\rangle$ (a) , $\sigma$ (b)  and
of the ratio of the total number of clusters at time $t$ to the total
initial number of clusters, $N(t)/N(0)$ (c) for initial log-normal
GCMF with $\langle \log M\rangle_i=4.5$ and 
$\sigma_i=1$ (circles) and with $\langle \log M\rangle_i=6$ and
$\sigma_i=1$ (triangles).\\
Figure 13  Trajectories in the plane $\sigma$-$\langle \log M\rangle$ (see
sect.3.4 for a detailed comment on the figure)\\
Figure 14 Contour plot in the plane $\sigma_i-\langle \log M\rangle_i$ of the
ratio, $F_N$, of the total number of clusters surviving after 15 Gyr
to the total 
initial number of clusters (a), and of the ratio,$F_M$,  of the total
mass of all 
surviving clusters after 15 Gyr to the total initial mass of all clusters
(b).\\ 
Figure 15 Contour plot in the plane $\sigma_i-\langle \log M\rangle_i$ of the
current value of cluster disruption rate (see text for definition).\\
Figure 16 Contour plot of the core radius, $R_c$, of the spatial
distribution (at $t=15$ Gyr) of clusters in the Galaxy
in the plane $\sigma_i-\langle \log M\rangle_i$.\\
Figure 17 Final values of $\langle \log M\rangle$ (a), $\sigma$ (b), $F_N$
(c), $F_D$ (d) from the simulations without the effects of disk shocking
versus those obtained taking into account disk shocking.\\
Figure 18 Initial and final GCMF for a simulation starting with a
power-law GCMF with $\alpha=2$ and $ M_{low}=10^4 M_{\odot}$.
Dots show the real GCMF obtained at $t=15 $  Gyr while solid
line shows the gaussian distribution with mean value and dipersion estimated
from the sample of surviving clusters as discussed in sect.2. Dashed
line shows the initial GCMF.\\
Figure 19 Final values of $\langle \log M\rangle$ (a) and $\sigma$ (b) versus
the low-mass cutoff in the initial GCMF for systems with a power-law initial
GCMF, $f(M) \propto M^{-\alpha}$. $\alpha=2$ (filled dots),
$\alpha=1.7$ (triangles ), $\alpha=1.5$ (crosses). \\
Figure 20 Histogram of the ratio of the final to initial mass for
clusters surviving after one Hubble time in a system with an initial
log-normal GCMF with ($\mm_i=5.0,~\sigma_i=0.7$). ($N(15)$ is the
total number of clusters survived at $t=15$ Gyr)\\
Figure 21 (a) Histogram of $\log [t_d(t_H)/t_H]$ (where $t_d(t_H)$ is
the disruption timescale of a cluster at $t_H=10$ Gyr) for the sample
of clusters surviving at $t=t_H$ from an initial population with a
log-normal GCMF (with $\sigma_i=0.7$ and $\mm_i=5.0$). The dashed line
is the fit to the histogram assuming the initial distribution of
$t_d$ to be a power-law while the solid line is the resulting
distribution at $t=t_H$ adopting the real initial distribution of
$t_d$ (see panel (b)) which is a log-normal distribution. (b)
Histogram of the initial values of $\log t_d$ for the system of
clusters with an initial log-normal GCMF (with $\sigma_i=0.7$ and $\mm_i=5.0$).
The solid line shows the best fit log-normal distribution
($\sigma=0.67$, $\langle \log [t_d(0)/t_H] \rangle=0.15$).\\
Figure 22 Initial ($t=0$) and final (at $t=10$ Gyr) distribution of
$\log t_d/t_H$ for three different systems: short dashed lines show
the initial (upper curve) and final (lower curve) distributions for a
system with an initial log-normal distribution of $t_d$ with
($\sigma=1.0$, $\langle \log [t_d(0)/t_H] \rangle=-0.7$); solid lines
(initial distribution is the upper curve and final distribution is the
lower curve) correspond to a  system  with an initial log-normal distribution
of $t_d$ with 
($\sigma=0.67$, $\langle \log [t_d(0)/t_H] \rangle=0.15$); long dashed
lines (initial distribution is the upper curve and final distribution is the
lower curve) correspond to a system with an initial power-law
distribution of $t_d$, $f_i \sim t_d^{-q}$, with $q=1.6$.\\
The normalizations of the initial distributions have been chosen so to
have the same total number of surviving clusters in the final samples
at $t=t_H$.
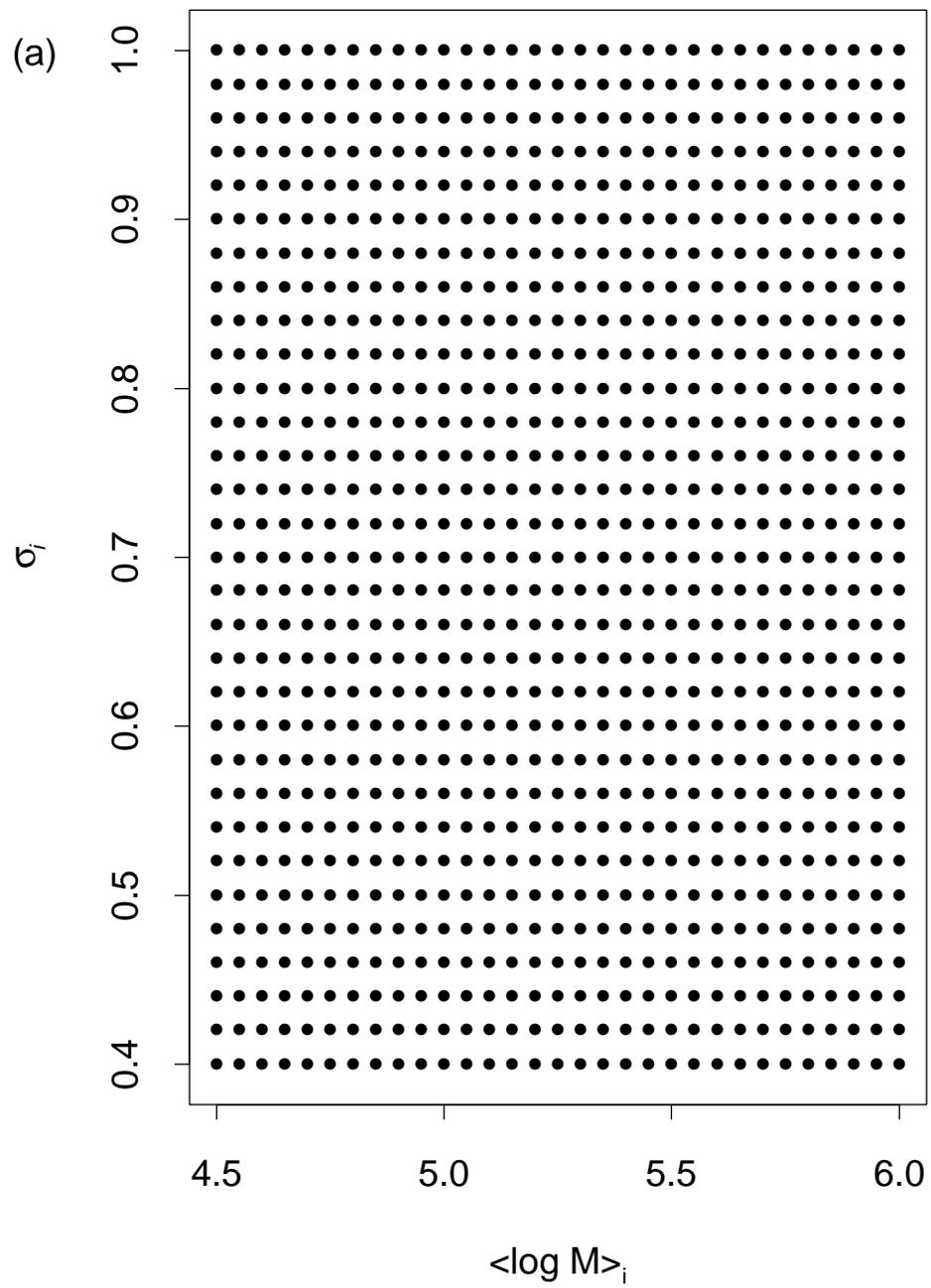
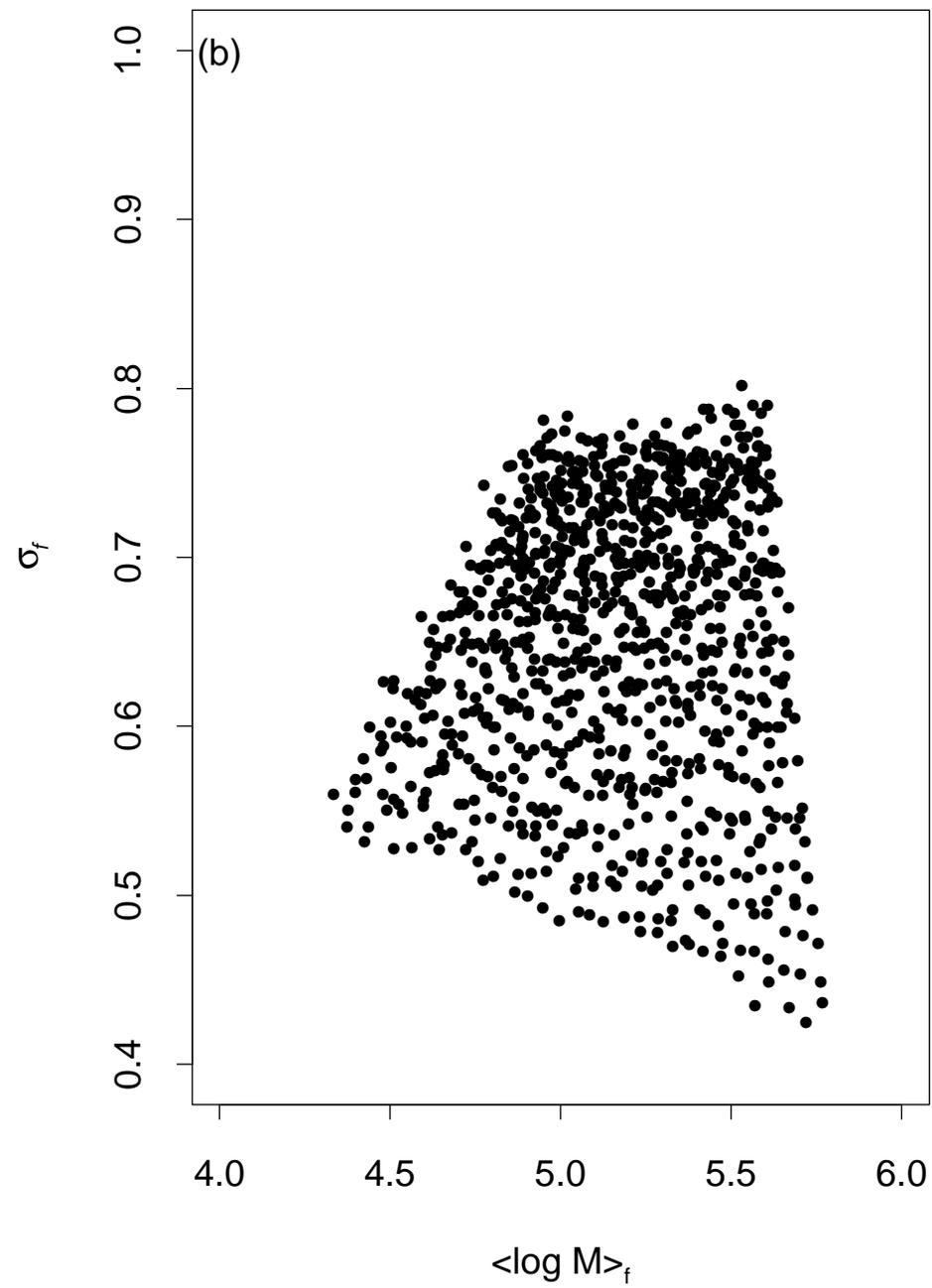
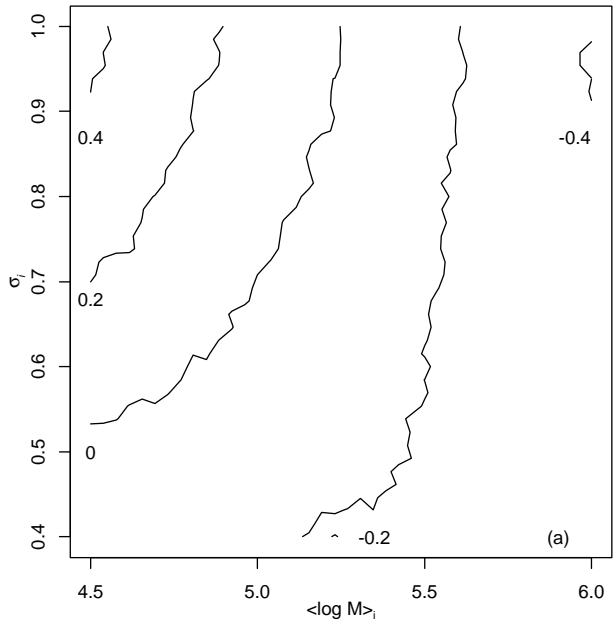
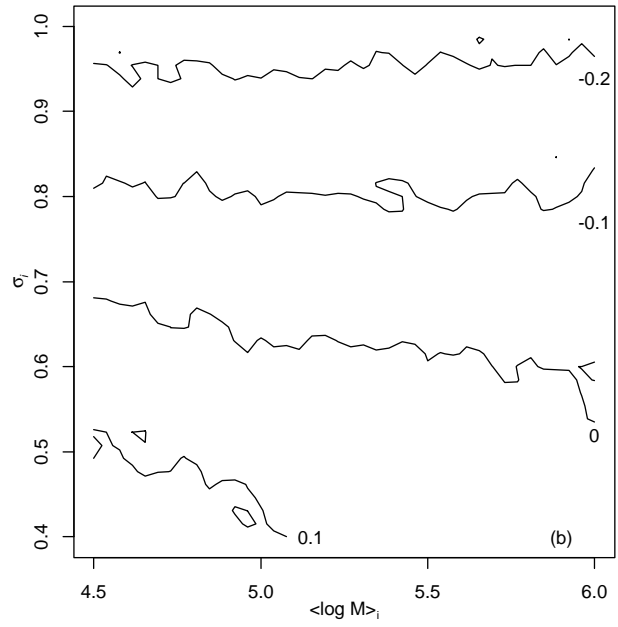
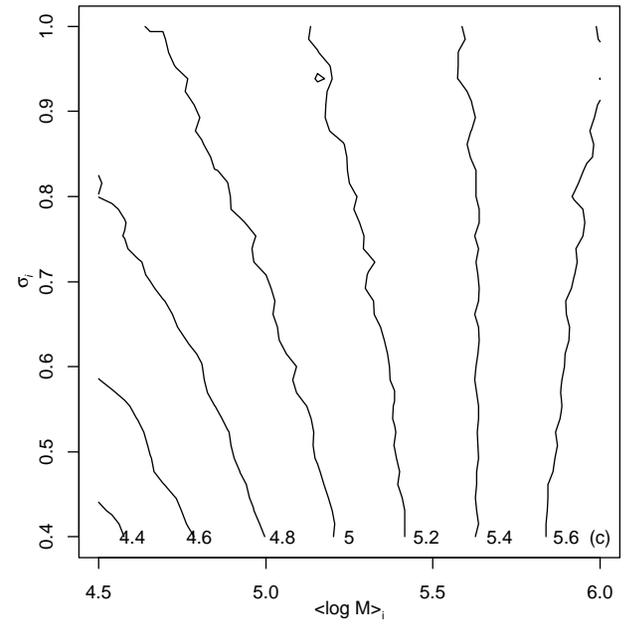
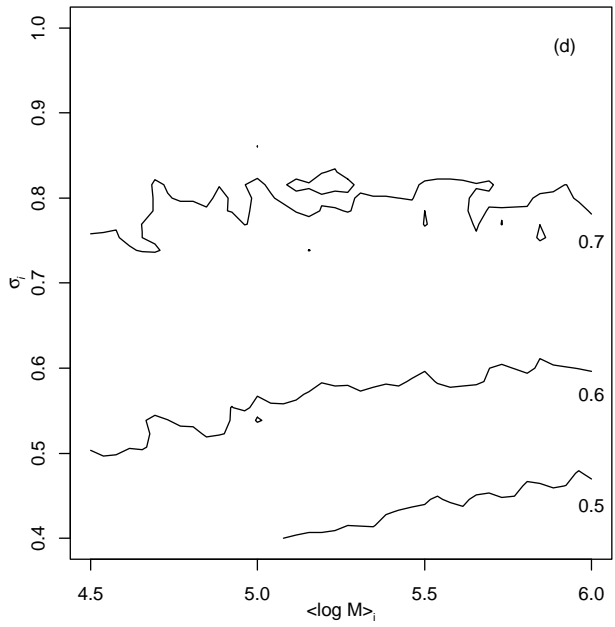
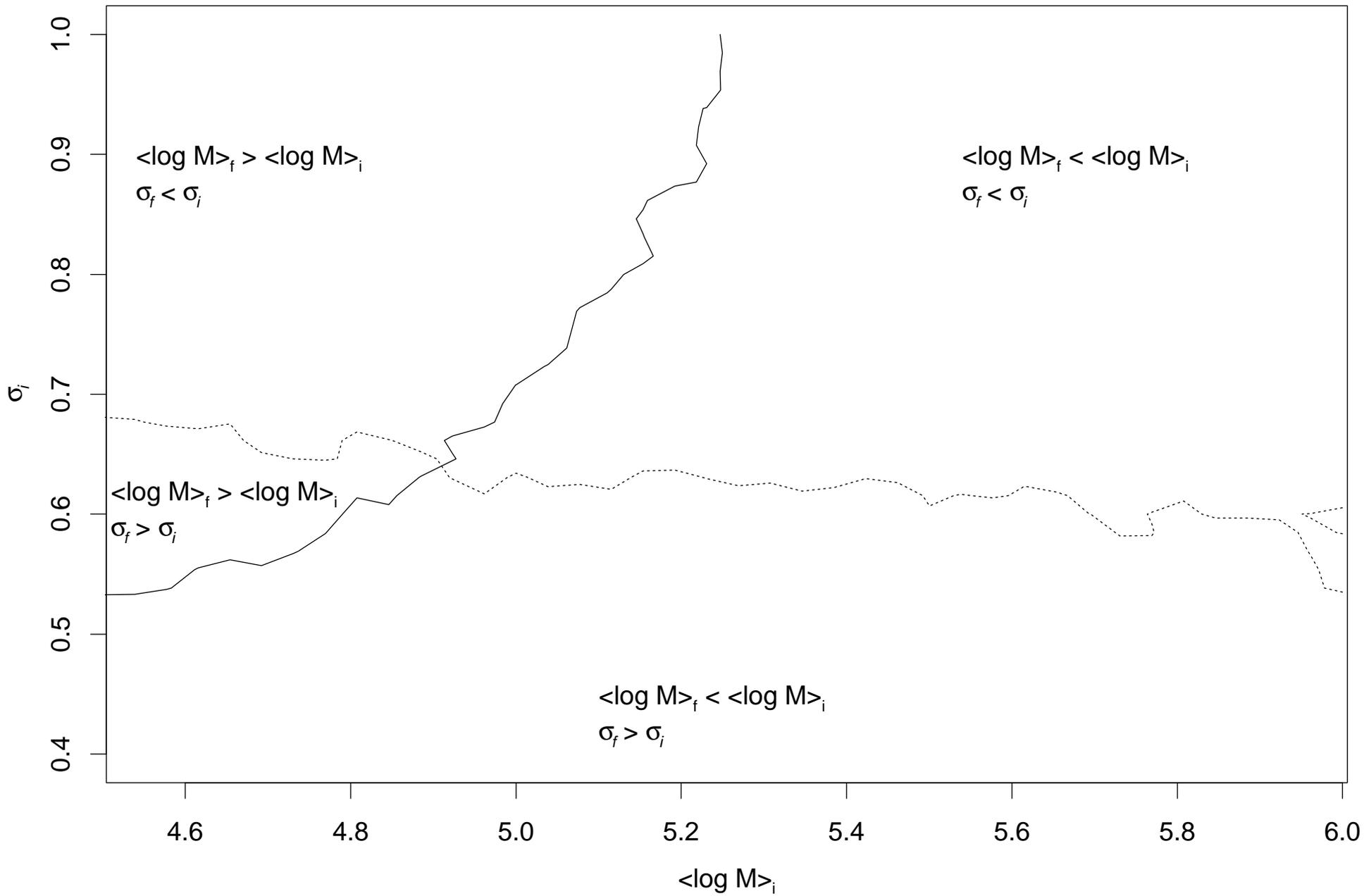
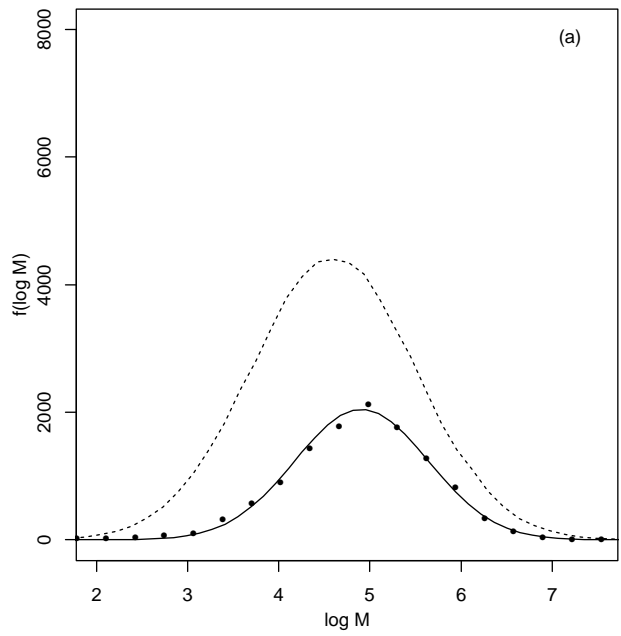
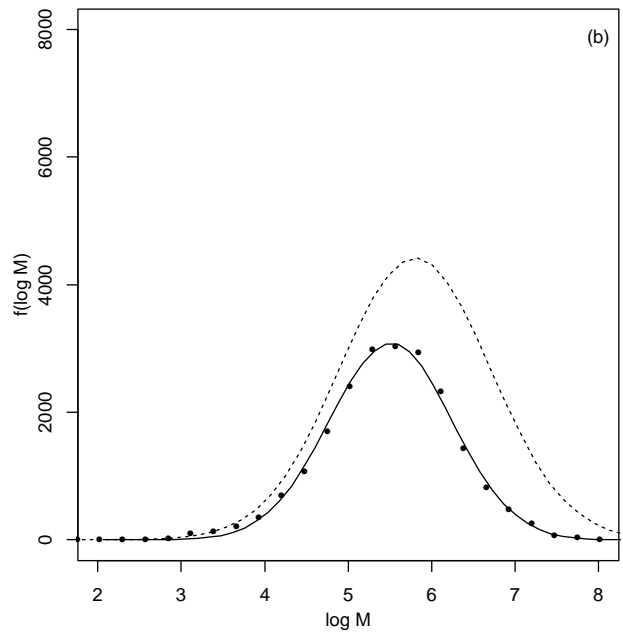
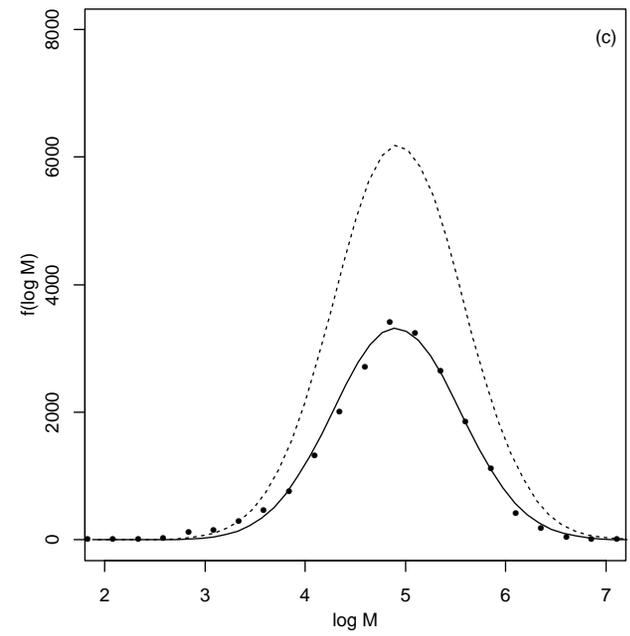
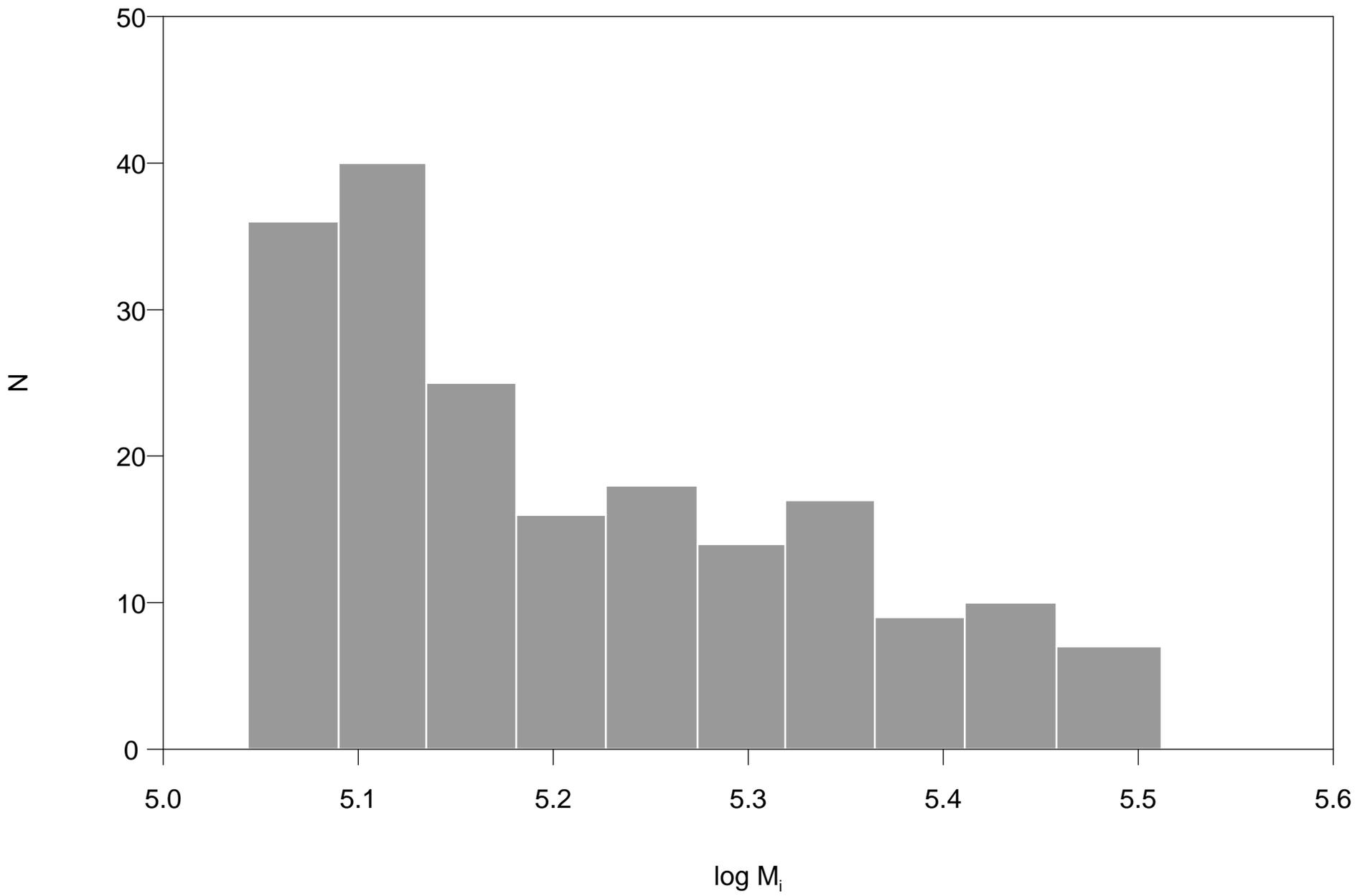
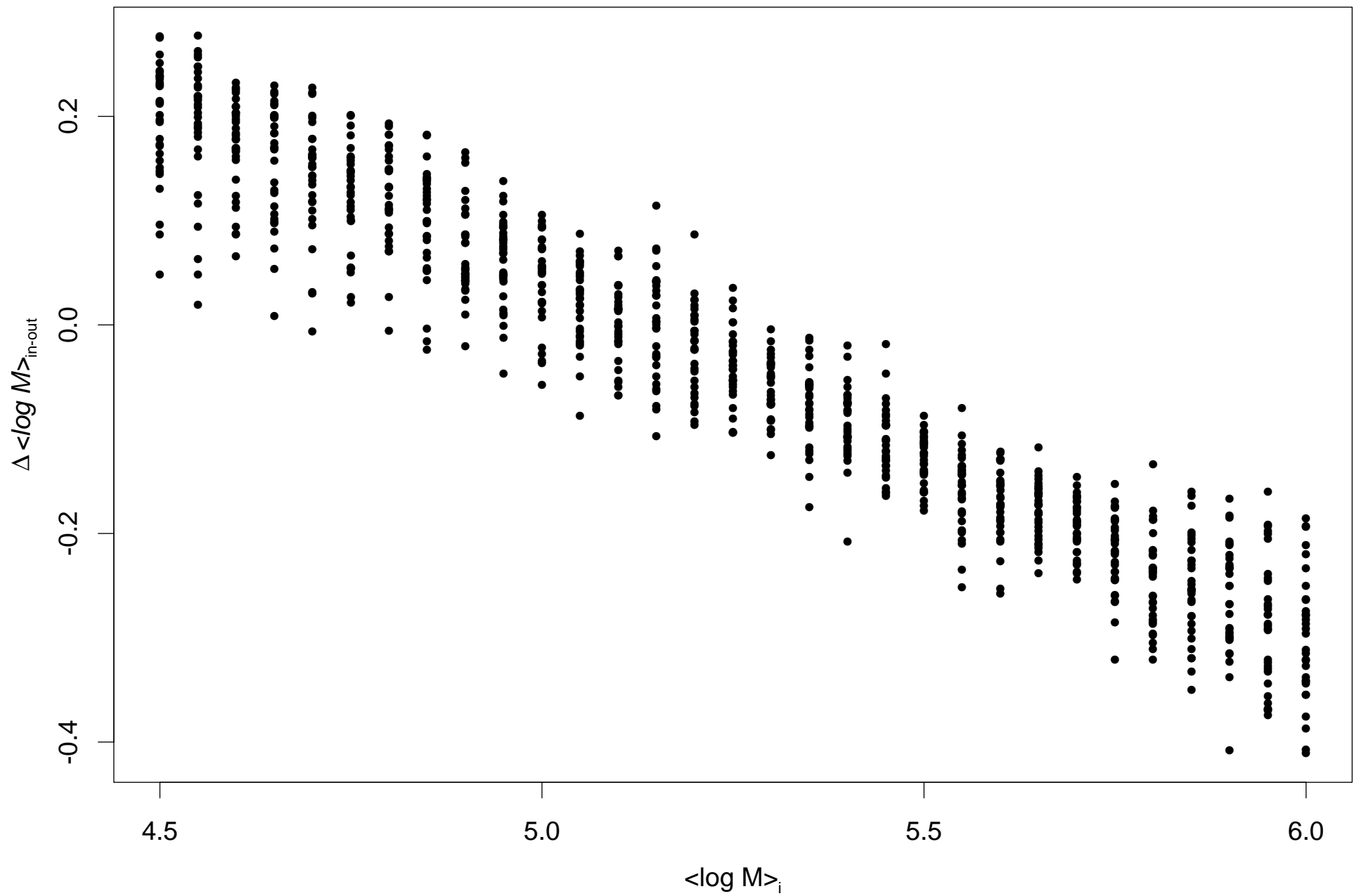
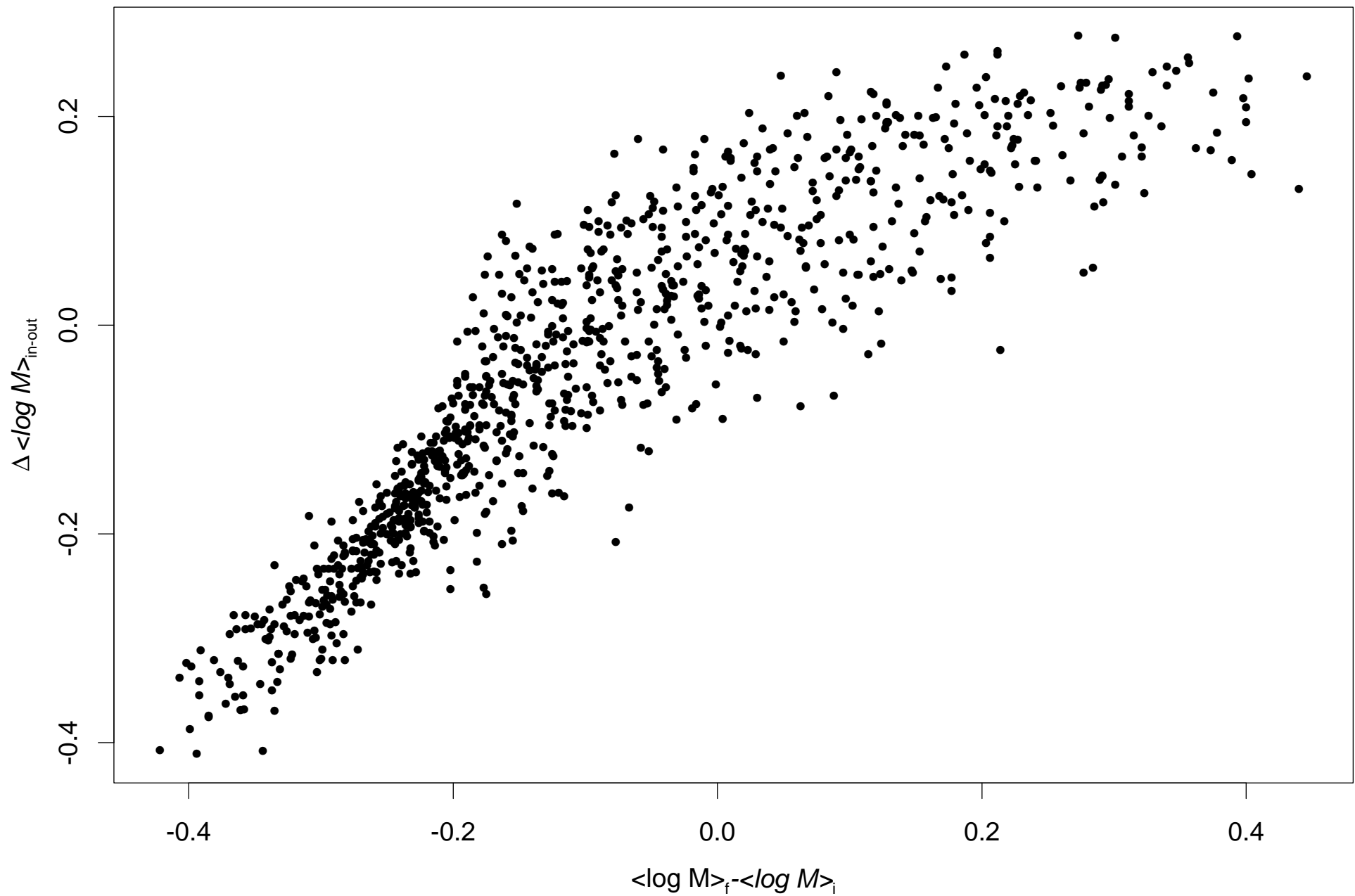
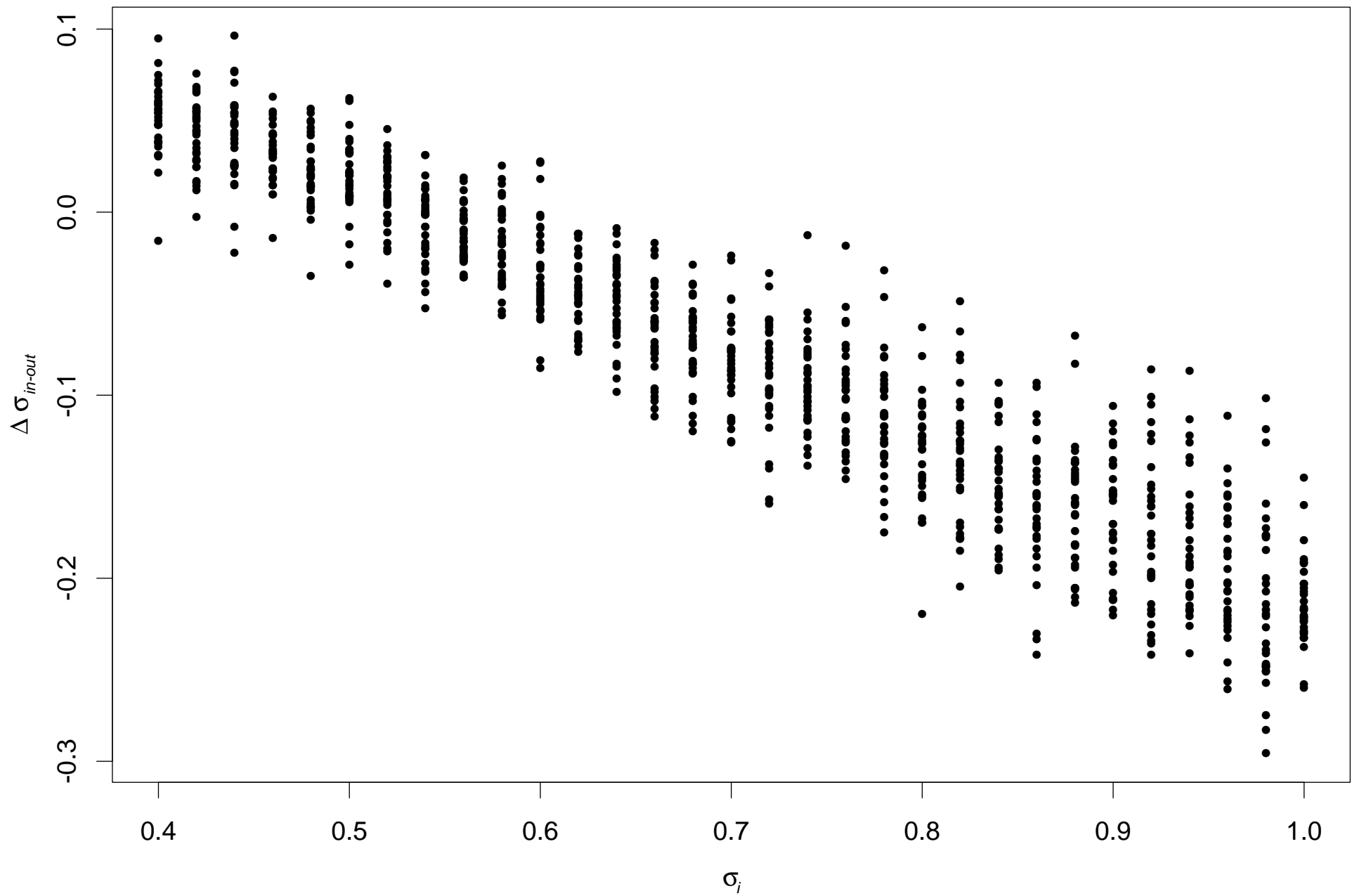
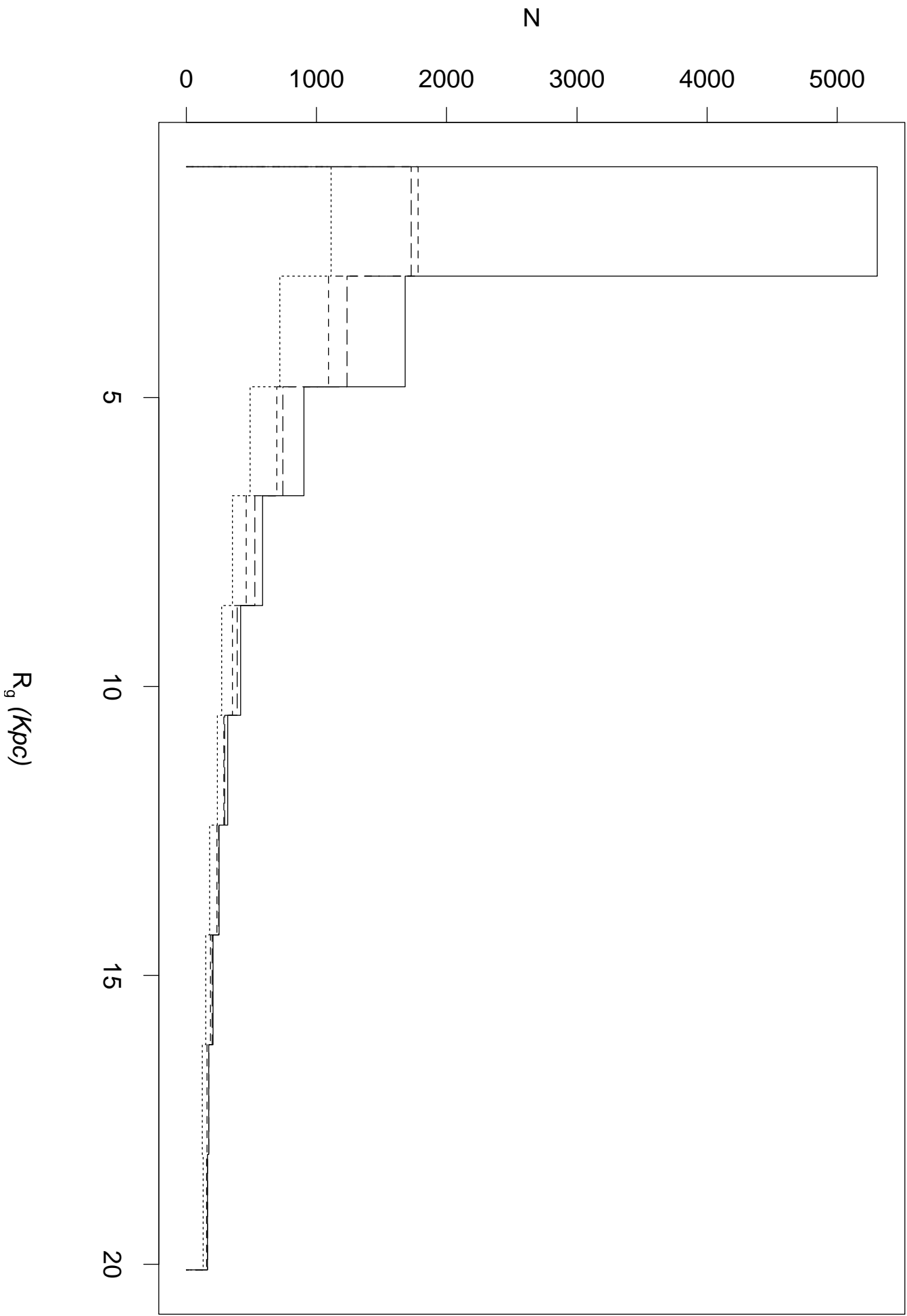
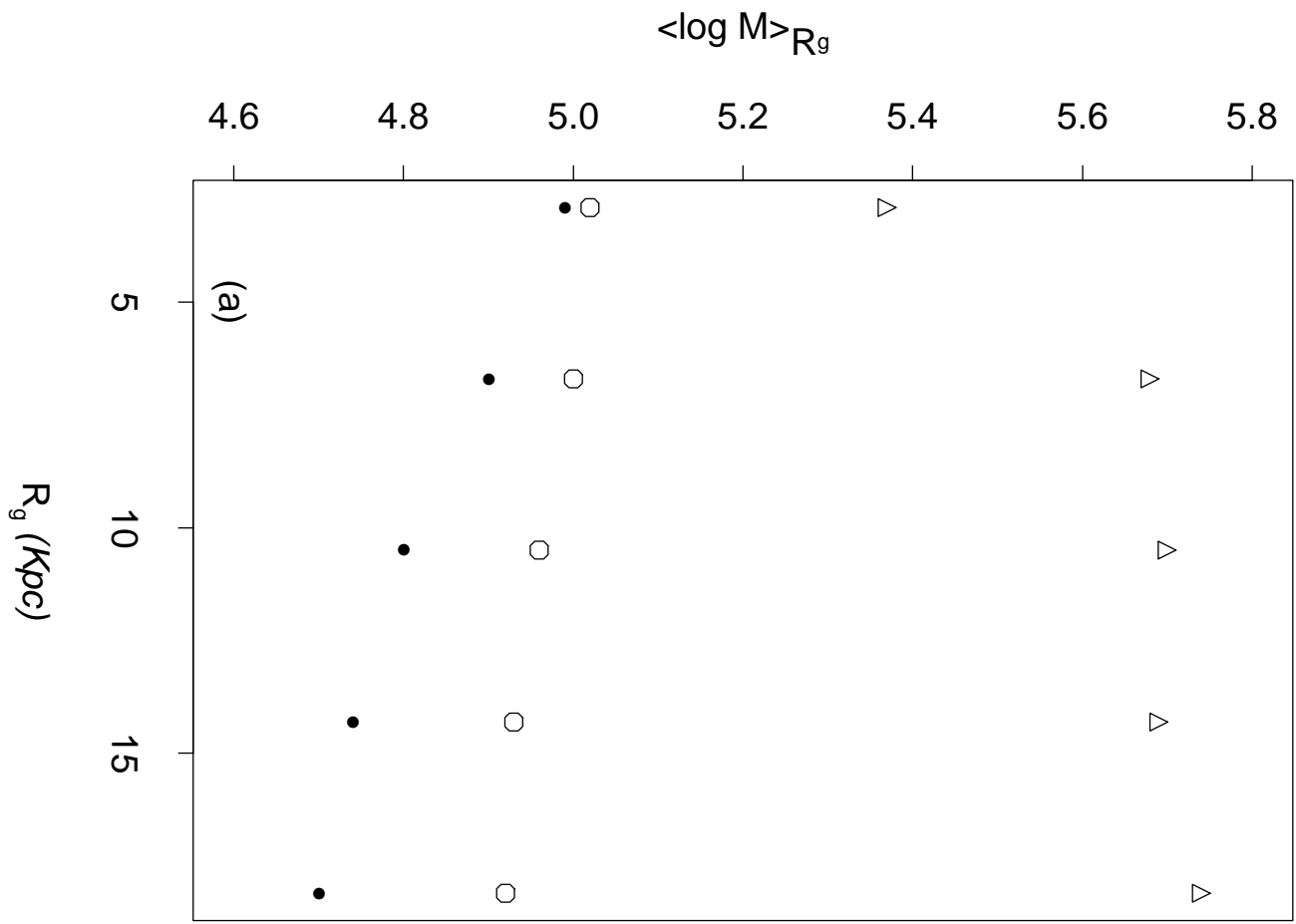
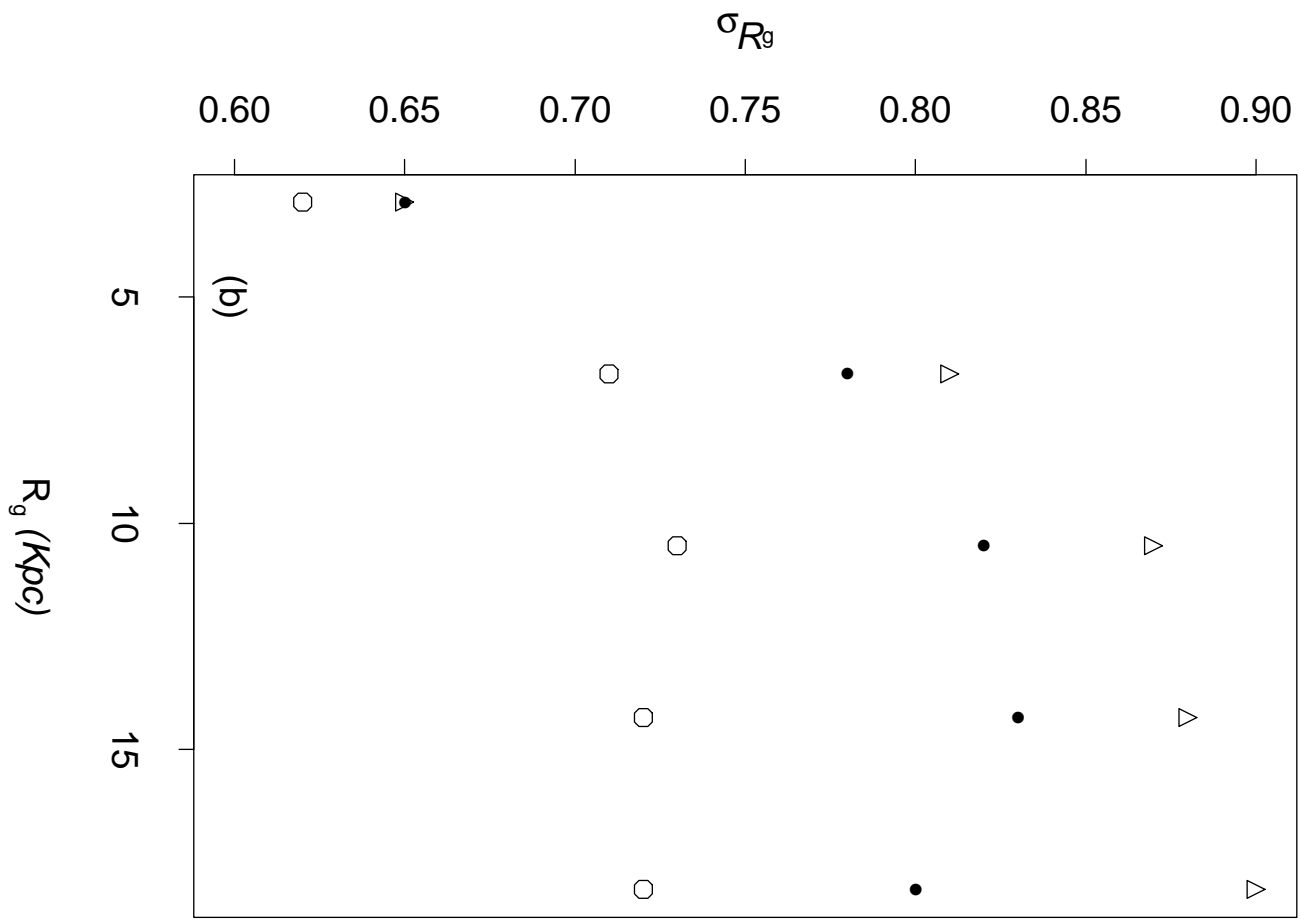
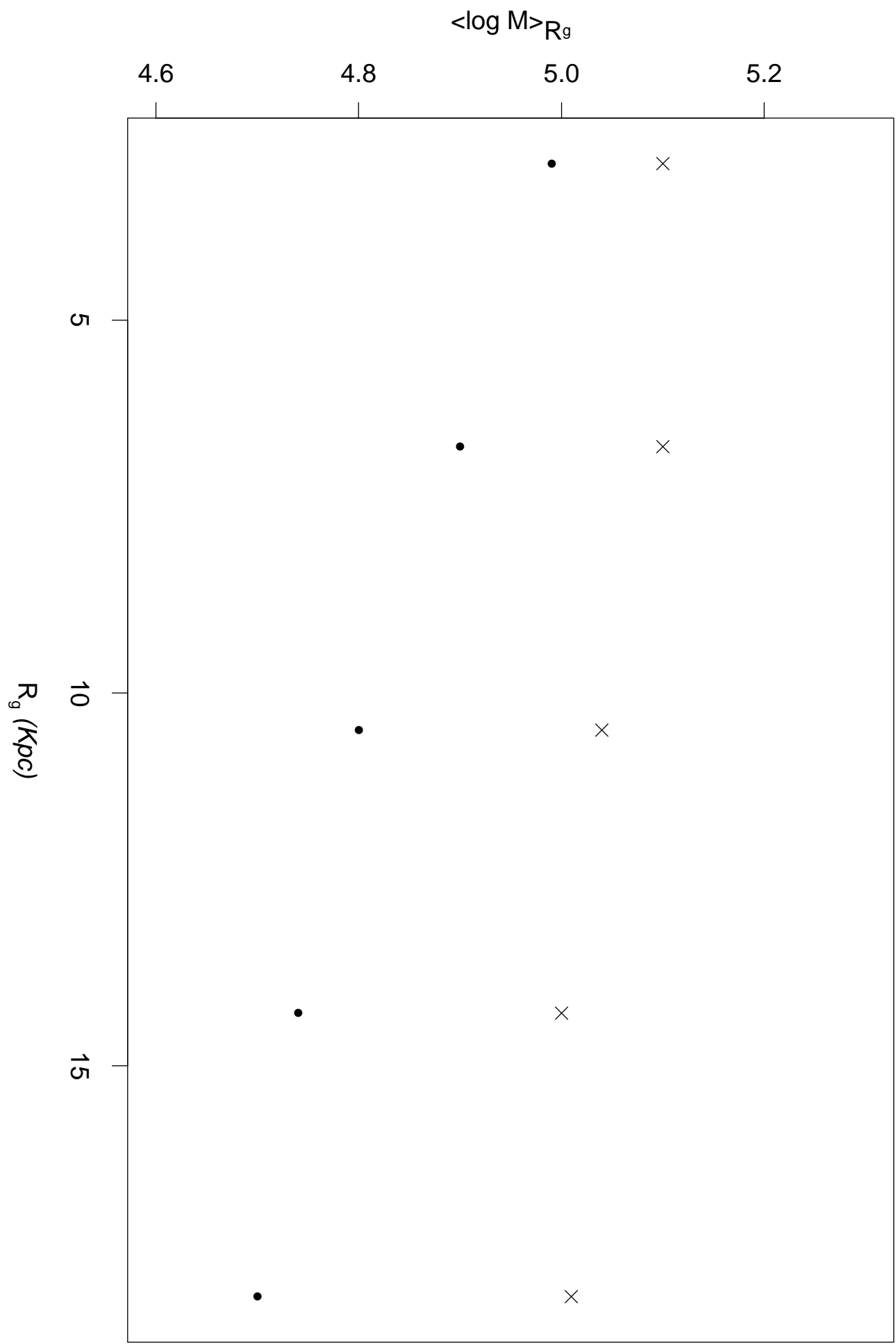
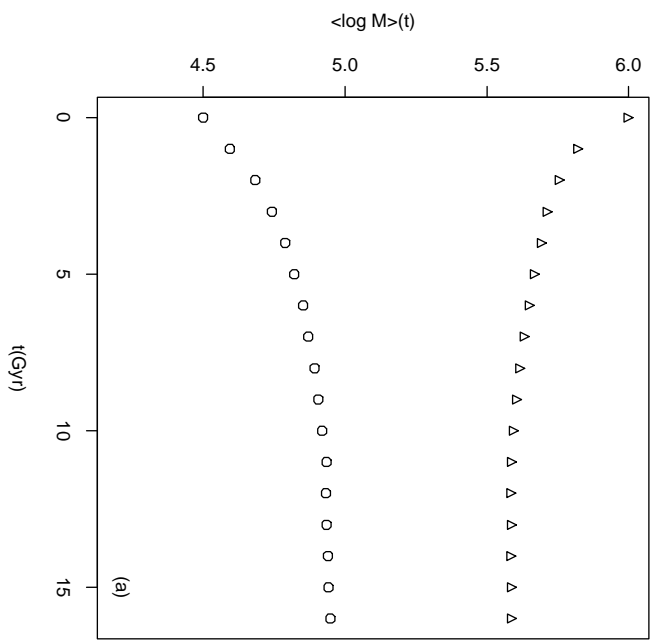
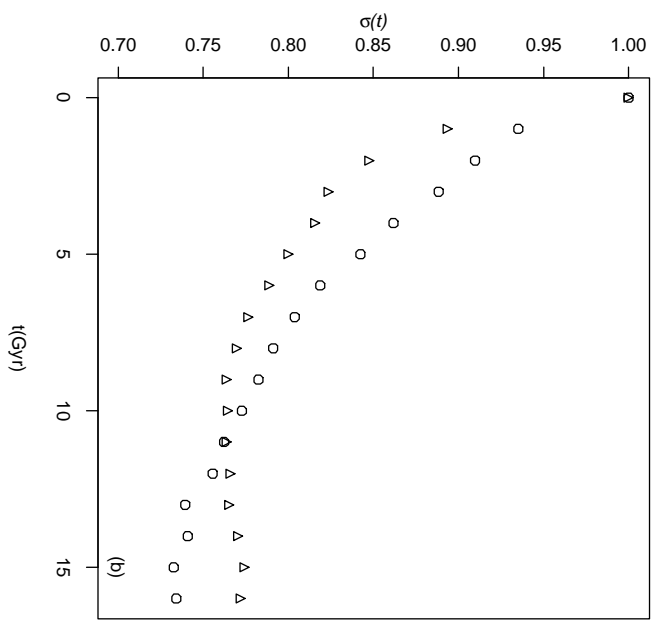
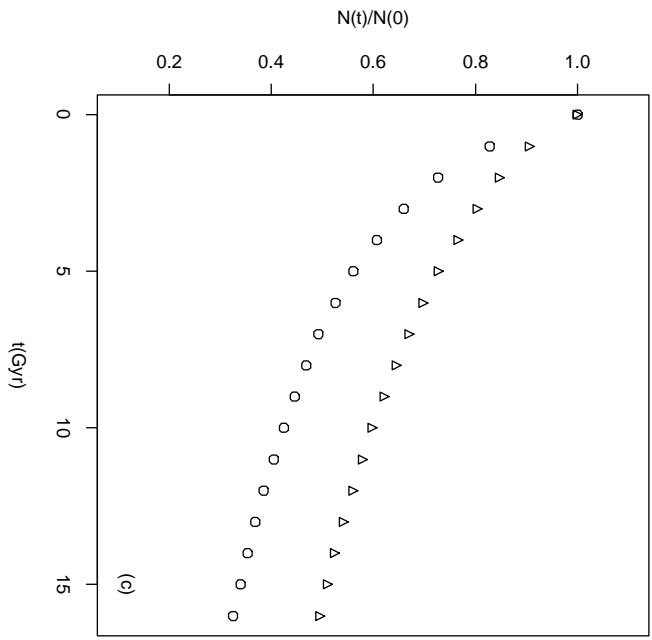
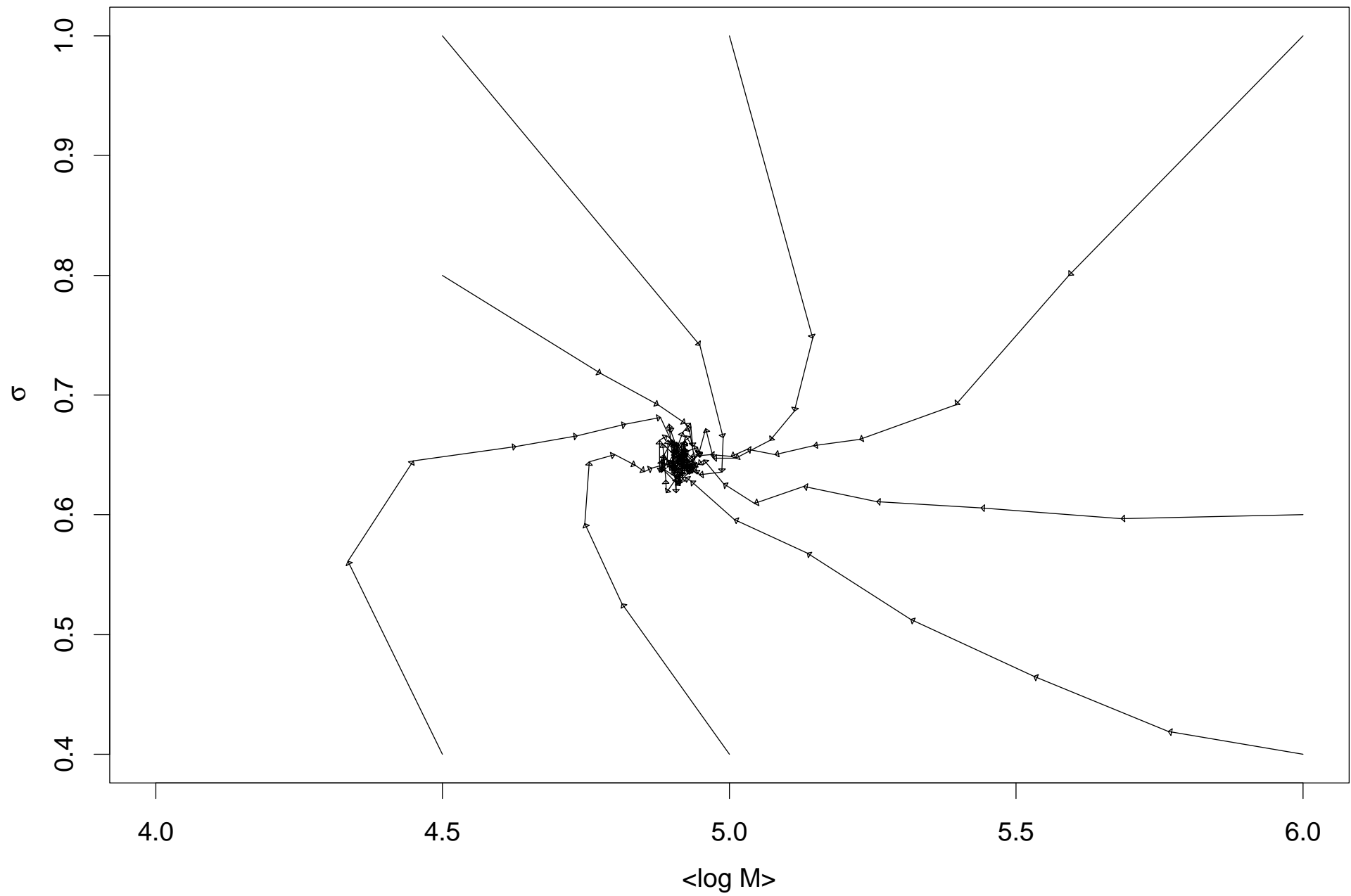
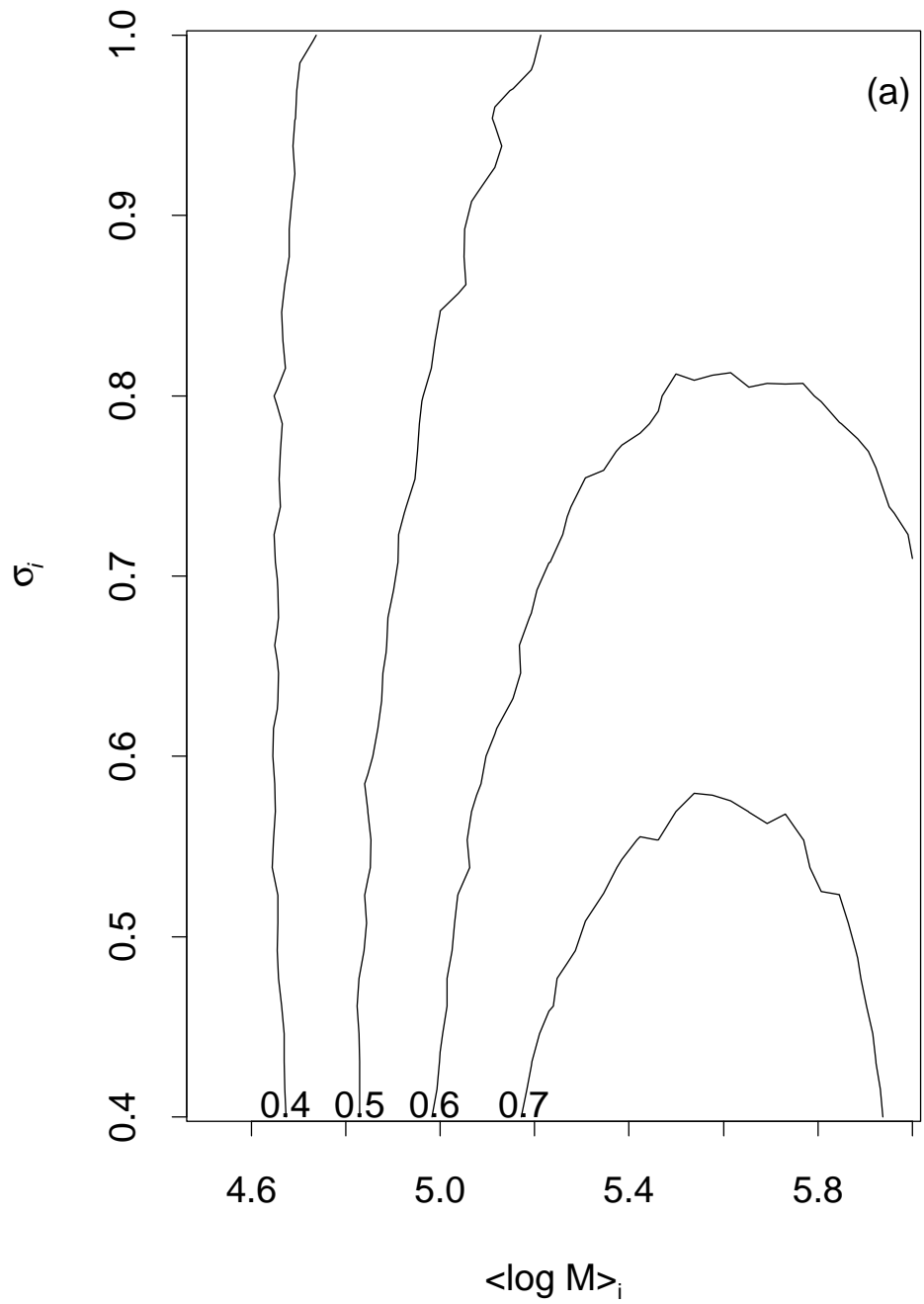
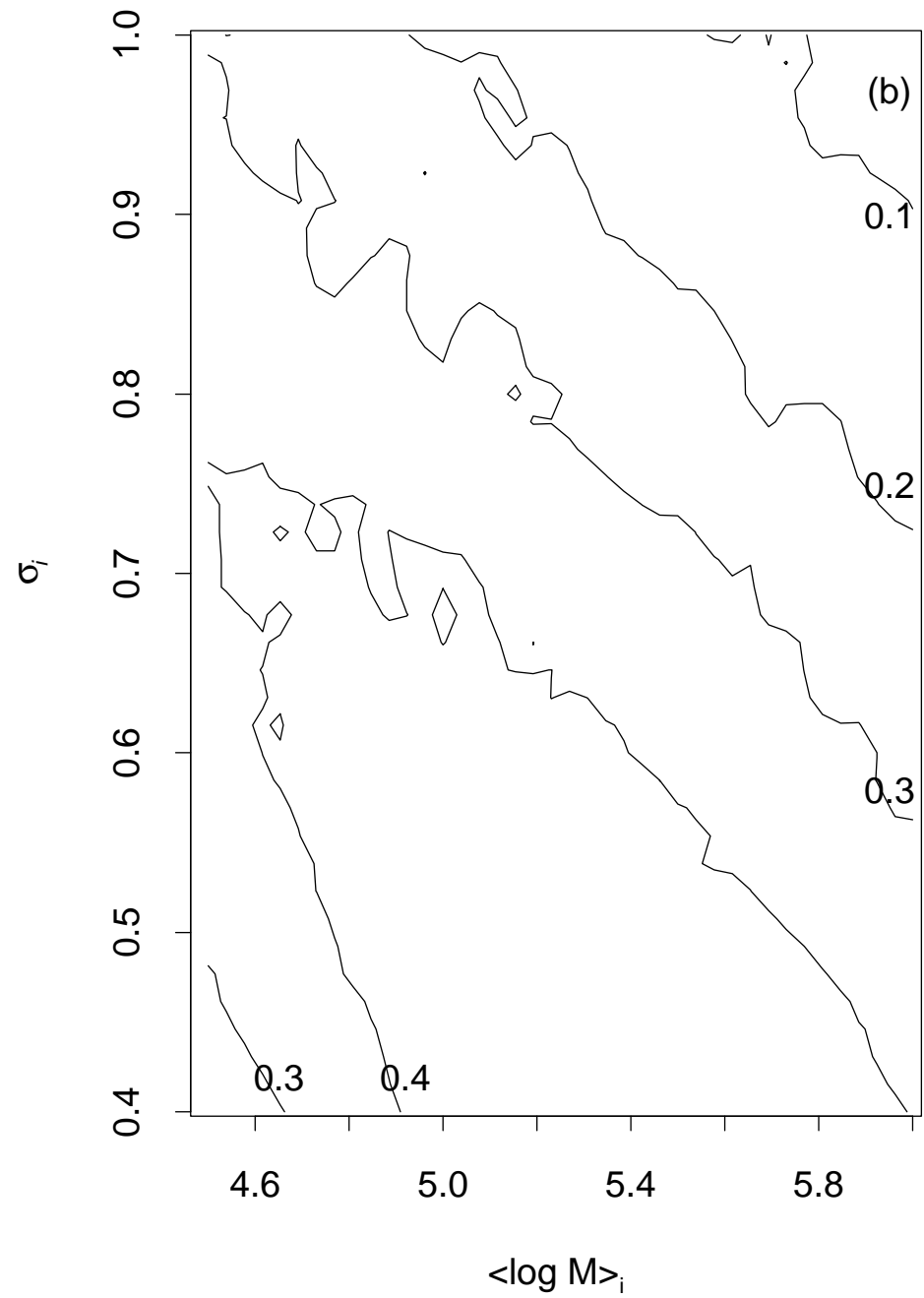
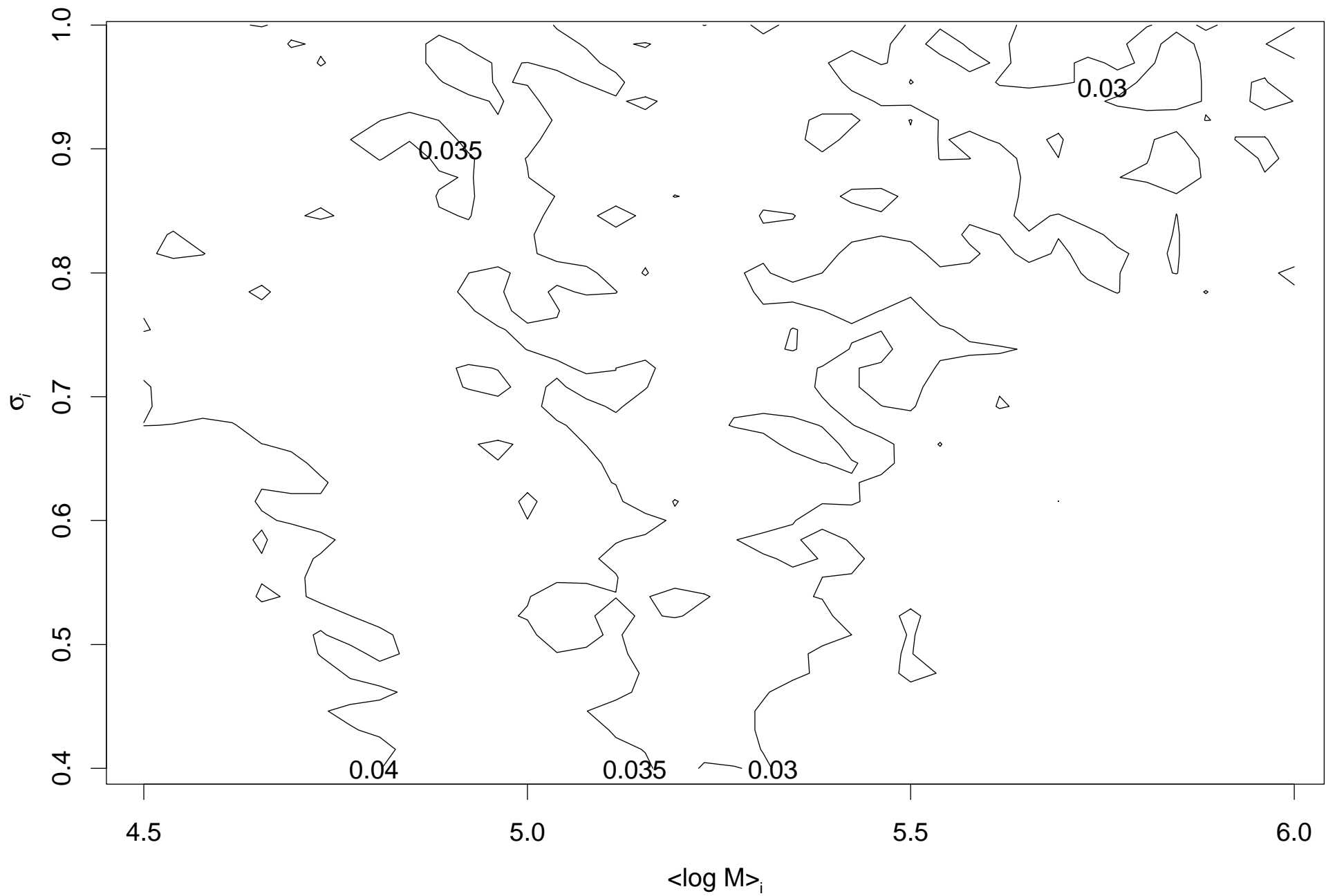
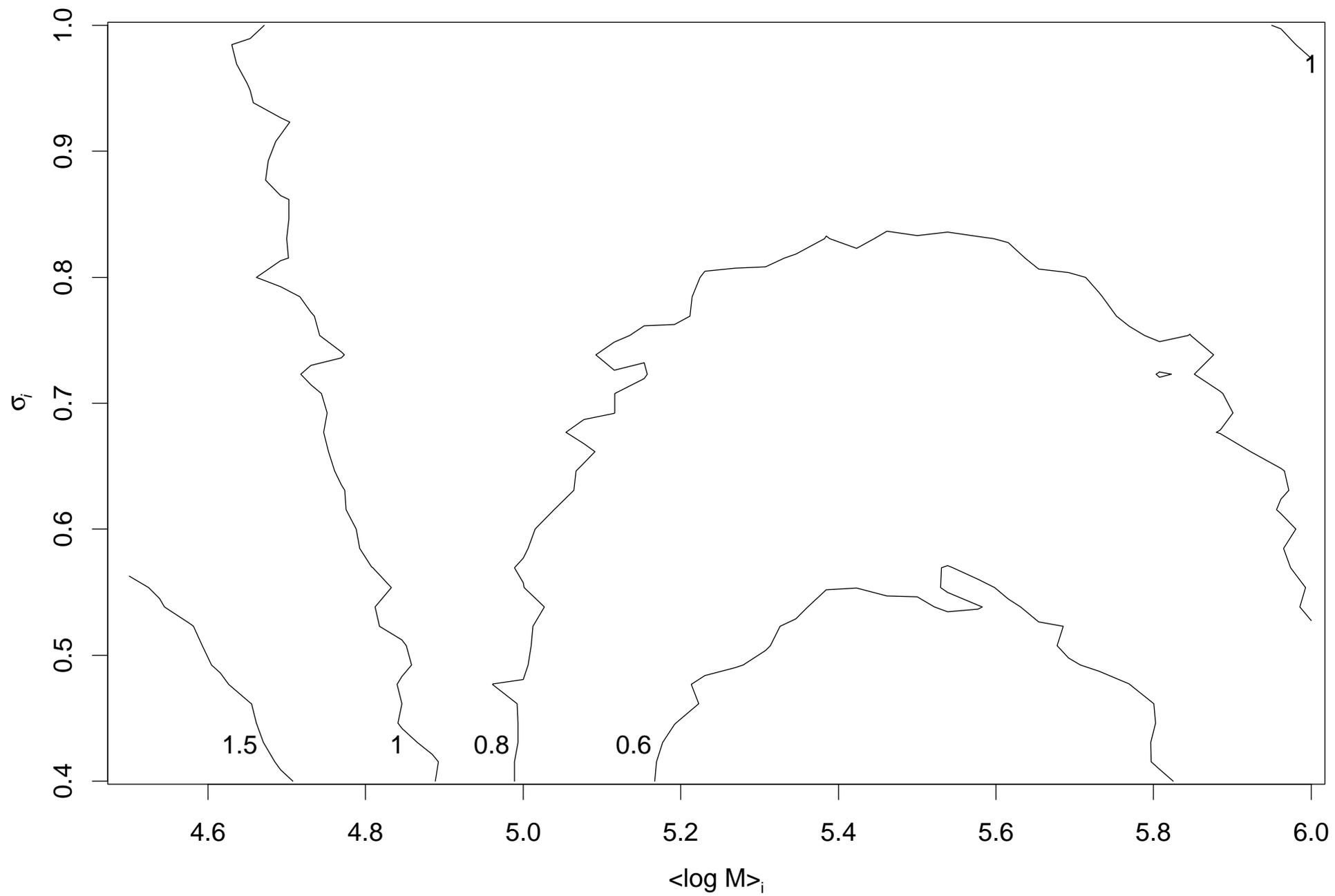
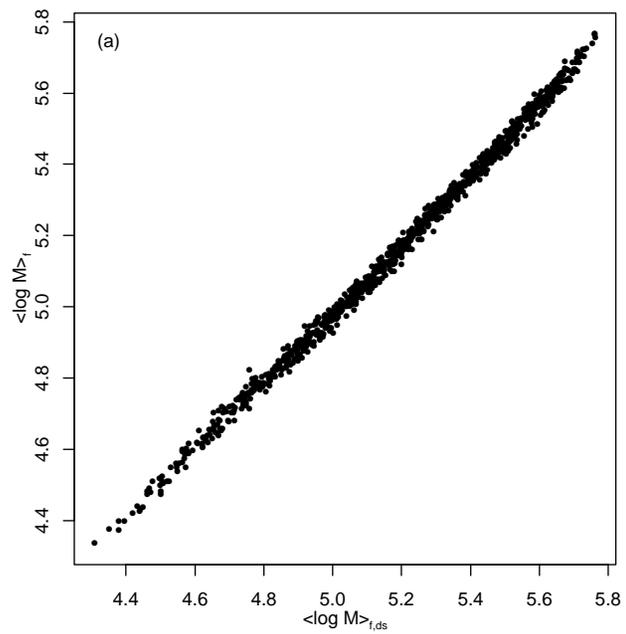
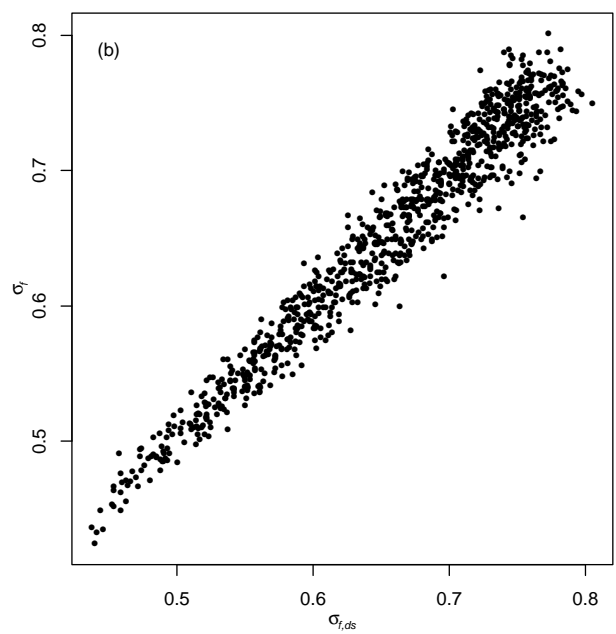
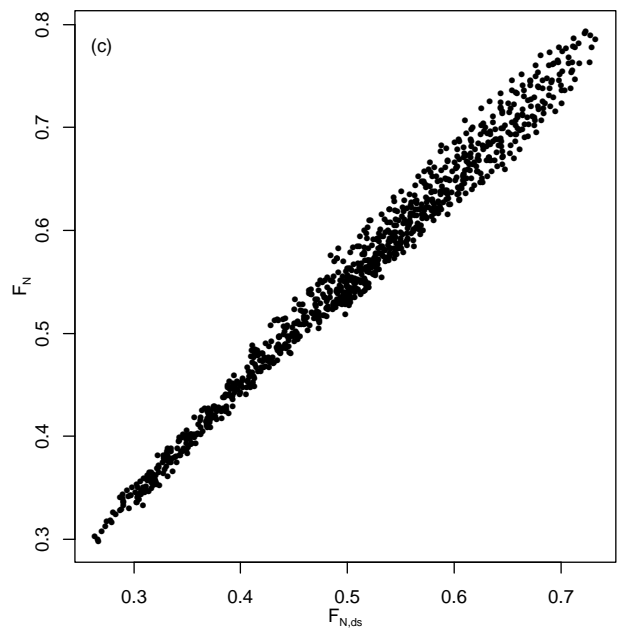
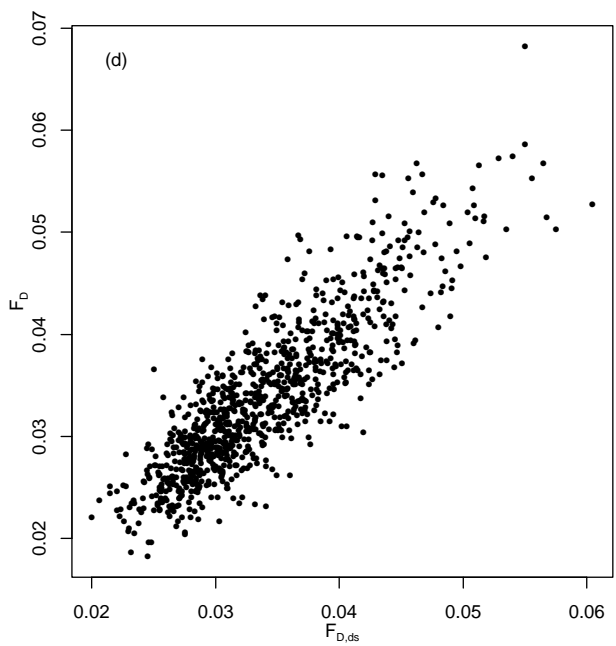
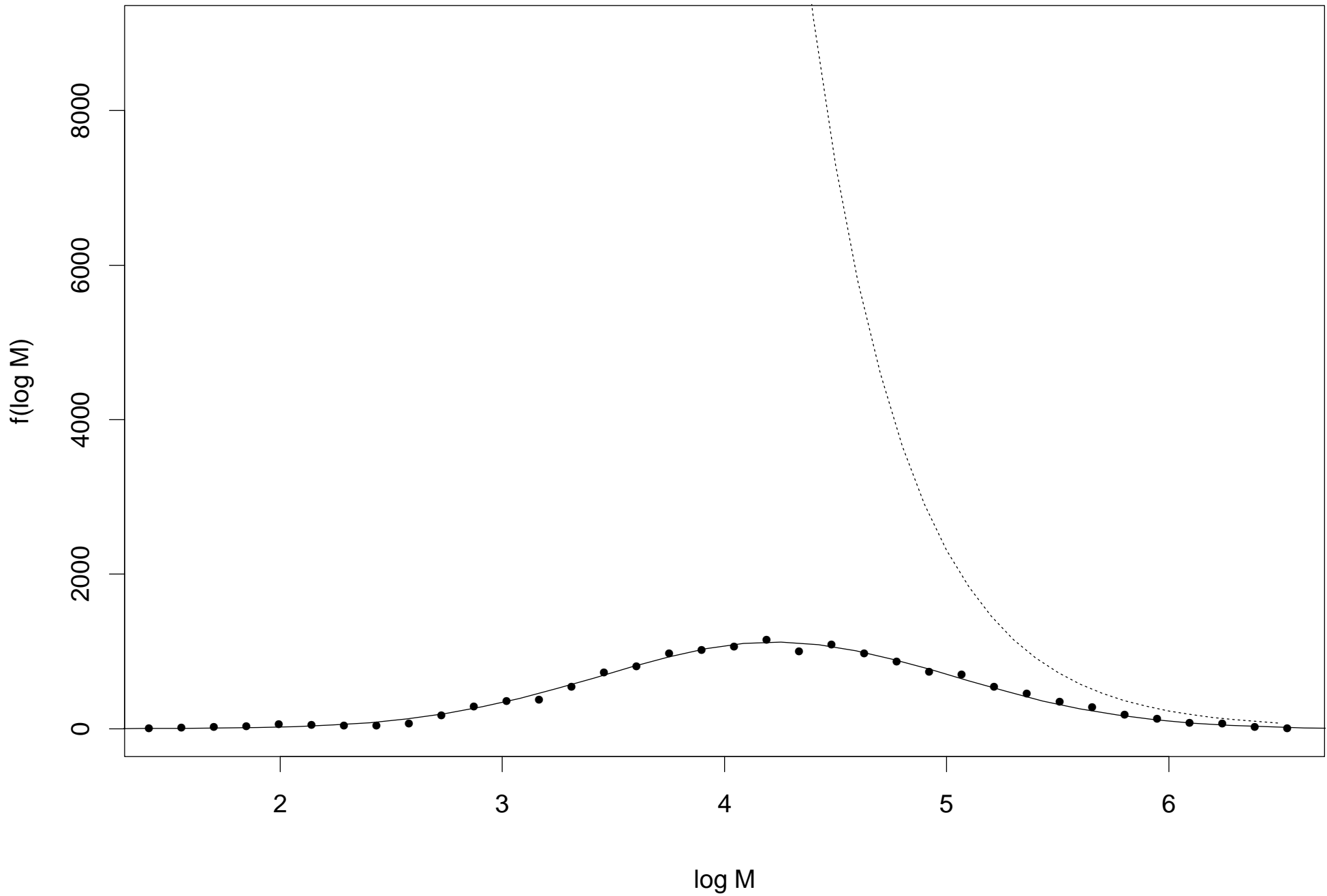
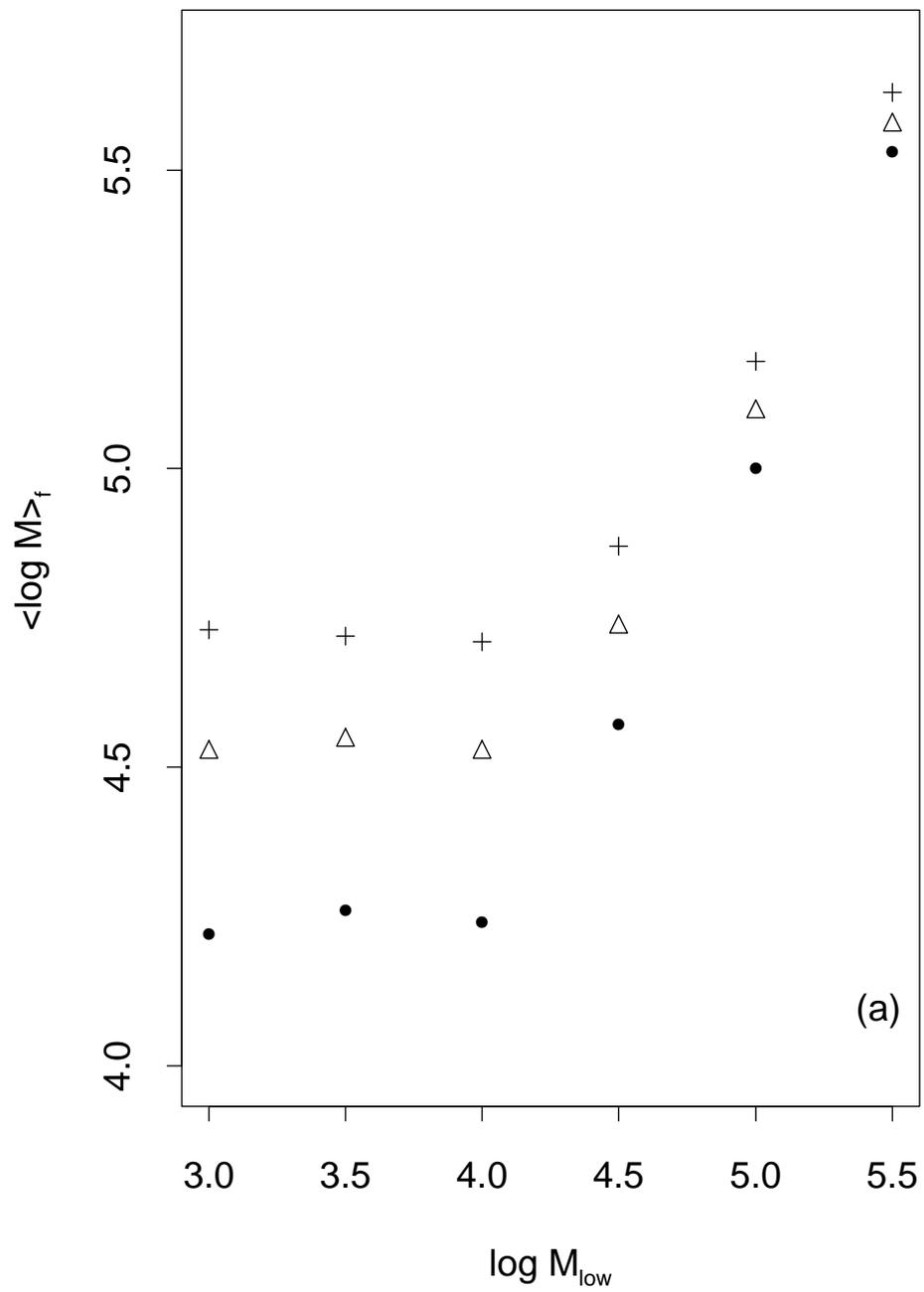
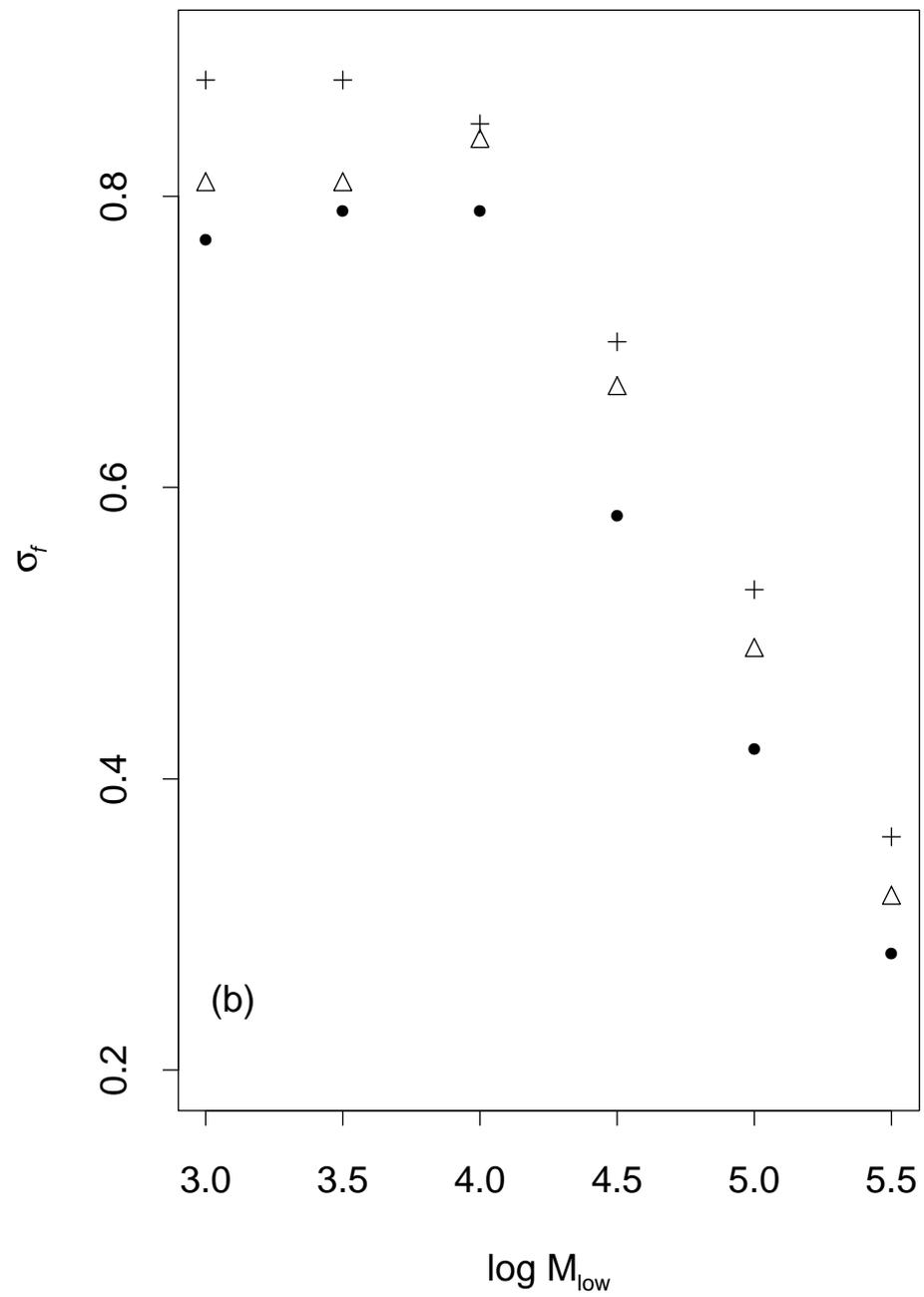
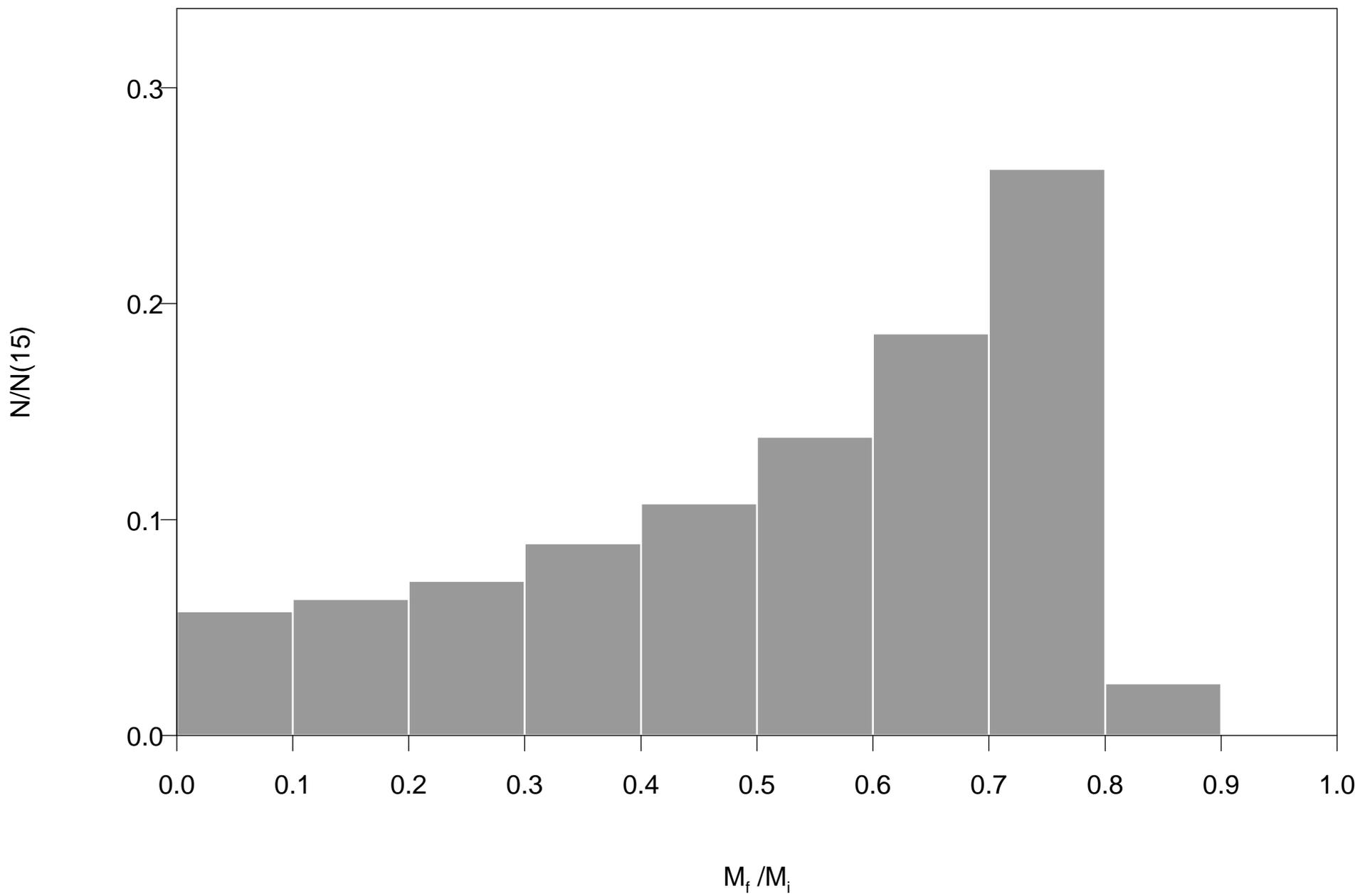
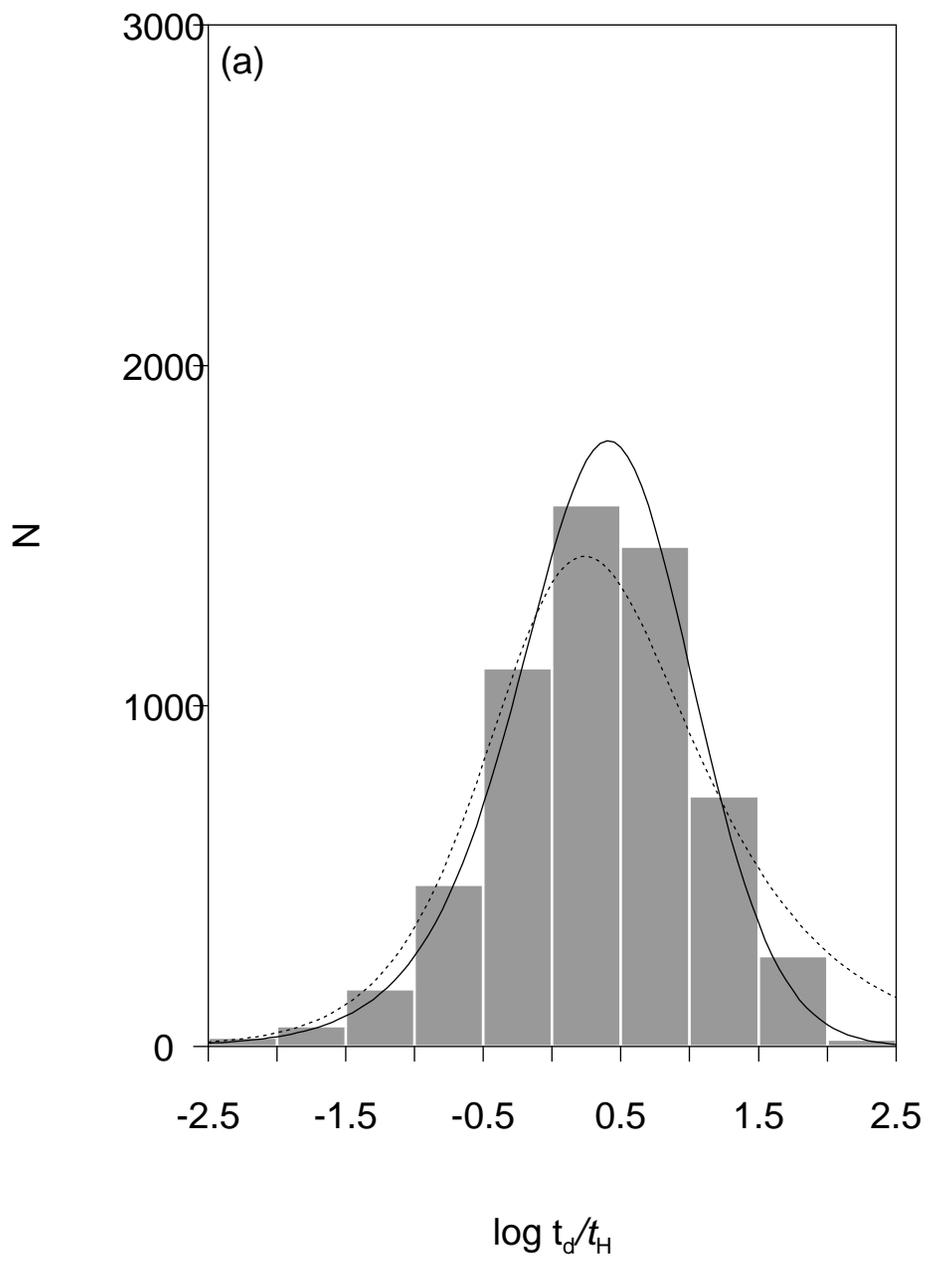
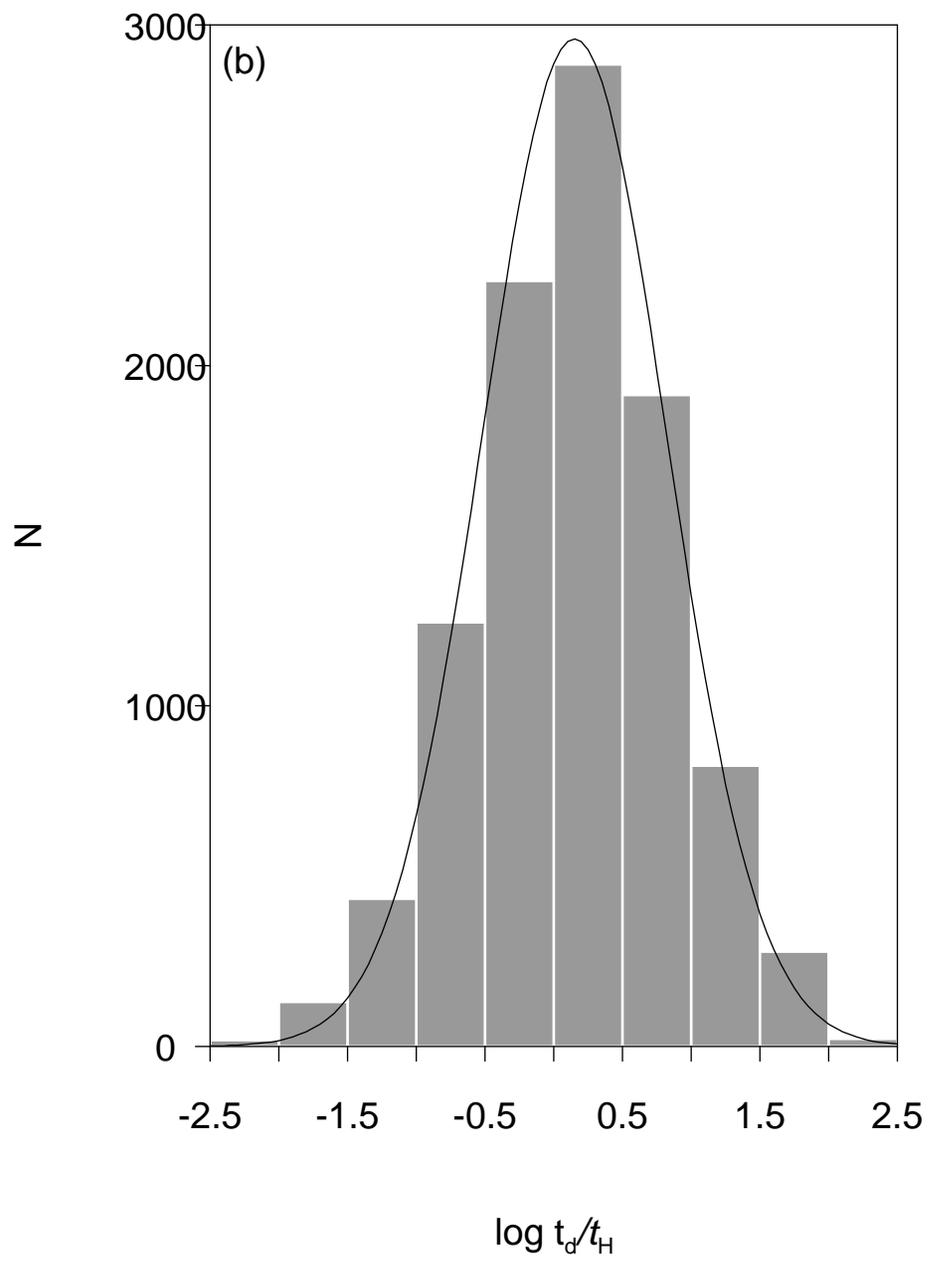
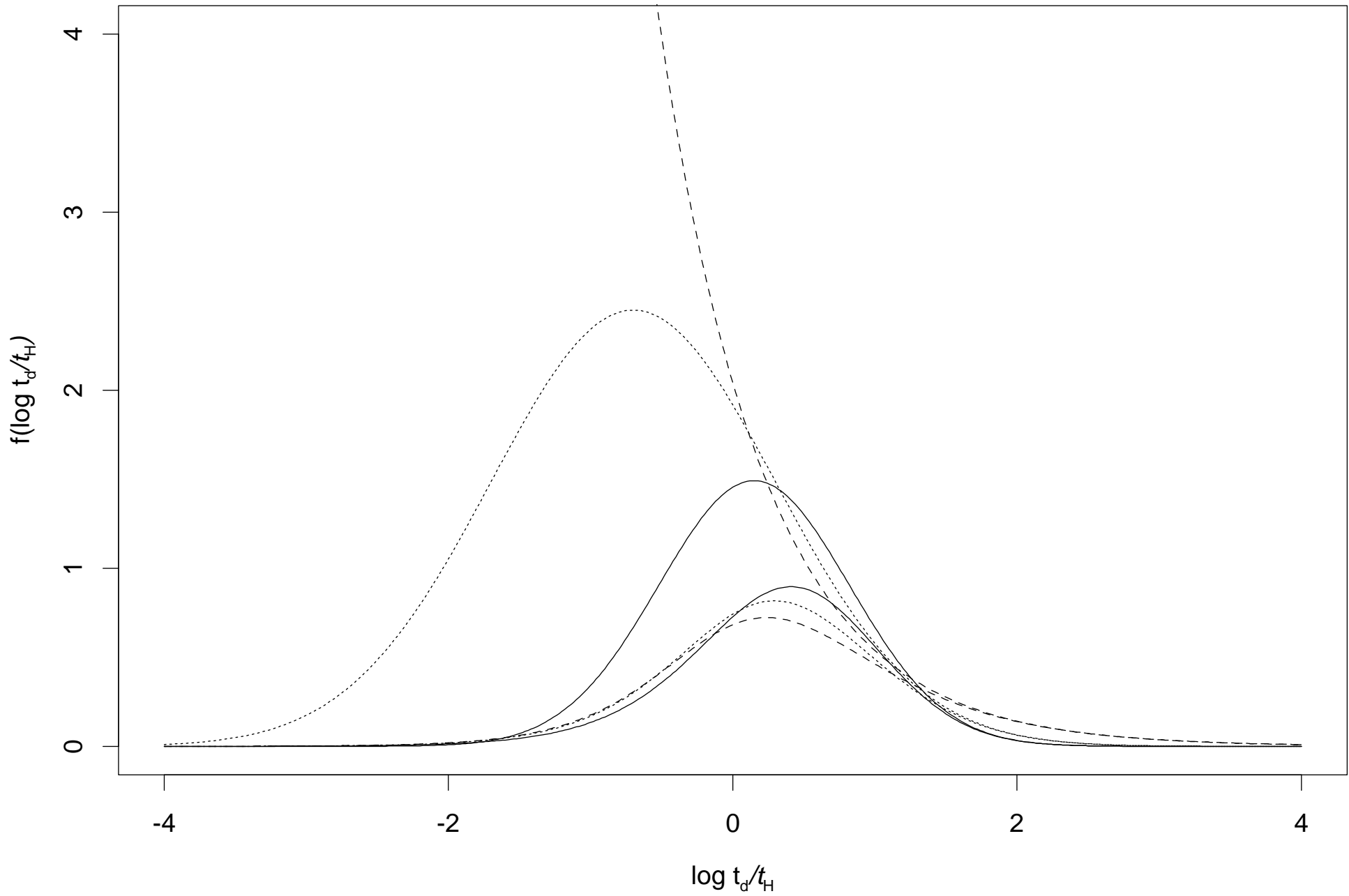
\end{document}